\documentclass[final]{siamltex}

\usepackage[dvips]{graphicx}
\usepackage[centertags]{amsmath}
\usepackage{amsfonts,amssymb,latexsym,epsfig}
\usepackage{graphicx}
\usepackage{dcolumn}
\usepackage{bm}
\usepackage{times}
\usepackage{enumerate}

\newcommand{\RR}{\mathbb R}
\newcommand{\bbE}{\mathbb E}
 
\newcommand{\bx}{\bar x}
\newcommand{\s}{\sigma}

\newcommand{\ri}{\rho_{\infty}}
\newcommand{\la}{\langle}
\newcommand{\ra}{\rangle}

\newcommand{\E}{\mathbb{E}}

\newcommand{\calN}{\mathcal{N}}
\newcommand{\calU}{\mathcal{U}}

\newcommand{\bbR}{\mathbb{R}}

\newcommand{\w}{\omega}

\newcommand{\bs}{\bar \sigma}
\newcommand{\bg}{\bar g}

\title{The Moment Map: Nonlinear Dynamics of Density Evolution Via a Few
Moments}

\author{D. Barkley\thanks{Mathematics Institute,
        University of Warwick, Coventry, CV4 7AL
        ({\tt barkley@maths.warwick.ac.uk}).}
        \and I.G.\ Kevrekidis\thanks{Department of Chemical Engineering, 
	  PACM and Mathematics,
	  Princeton University, Princeton, NJ, 08544
        ({\tt yannis@Princeton.edu}).}
        \and A.M.\ Stuart\thanks{Mathematics Institute,
        University of Warwick, Coventry, CV4 7AL
        ({\tt stuart@maths.warwick.ac.uk}).}}

\begin{document}

\maketitle

\begin{abstract}

We explore situations in which certain stochastic and high-dimensional
deterministic systems behave effectively as low-dimensional dynamical systems.
We define and study moment maps, maps on spaces of low-order moments of
evolving distributions, as a means of understanding equations-free multiscale
algorithms for these systems.  We demonstrate how nonlinearity arises in these
maps and how this results in the stabilization of metastable states.  Examples
are shown for a hierarchy of models, ranging from simple stochastic
differential equations to molecular dynamics simulations of a particle in
contact with a heat bath.

\end{abstract}

\begin{keywords} 
Equation-free, multiscale, moment map, metastable states. 
\end{keywords}

\begin{AMS}

\end{AMS}

\pagestyle{myheadings}
\thispagestyle{plain}
\markboth{D. BARKLEY, I.G.\ KEVREKIDIS AND A.M.\ STUART}{THE MOMENT MAP}


\section{Introduction}
\label{sec:intro}

An equation-free framework has recently been developed as a means of
computationally analyzing the dynamical behavior of a large class of complex,
multiscale dynamical systems.  The systems may be either stochastic or
deterministic with many of degrees of freedom and subject to random initial
data.  The key observation behind the equation-free framework is that in many
cases the quantities of interest are averages or low-order moments of evolving
distributions which are smooth in space and time and which evolve effectively
as closed, low-dimensional systems.  In effect the low-order moments evolve as
though they are governed by reduced closed equations, even though the reduced
equations are not analytically available.  Algorithms for performing
scientific computing tasks such as numerical integration, or bifurcation and
stability analysis of these unavailable reduced equations have been developed
(e.g. coarse projective integration, coarse Newton-GMRES, see
\cite{PNAS,GKT,Manifesto,ShortManifesto} and references therein).  These
algorithms are based on traditional, continuum numerical analysis, ``wrapped"
around direct fine-scale simulation.  The purpose of this paper is to
establish a mathematical framework for understanding the behavior of coarse
dynamics and coarse bifurcation methods on problems which exhibit metastable
behavior.

\begin{figure}
\centerline{
\includegraphics[width=2.5in]{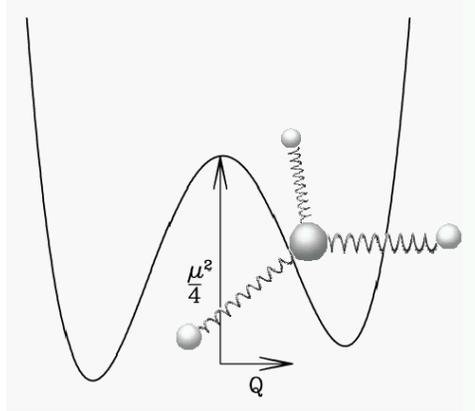}
}
\caption{Model system. A distinguished particle (light gray) sits in an
    asymmetric double well potential. The particle is coupled via
    linear springs to $N$ other (bath) particles (three are shown as
    illustration). The full system has $N+1$ degrees of freedom. }
\label{fig:sketch}
\end{figure}

\begin{figure}
\centerline{ \includegraphics[width=2.0in]{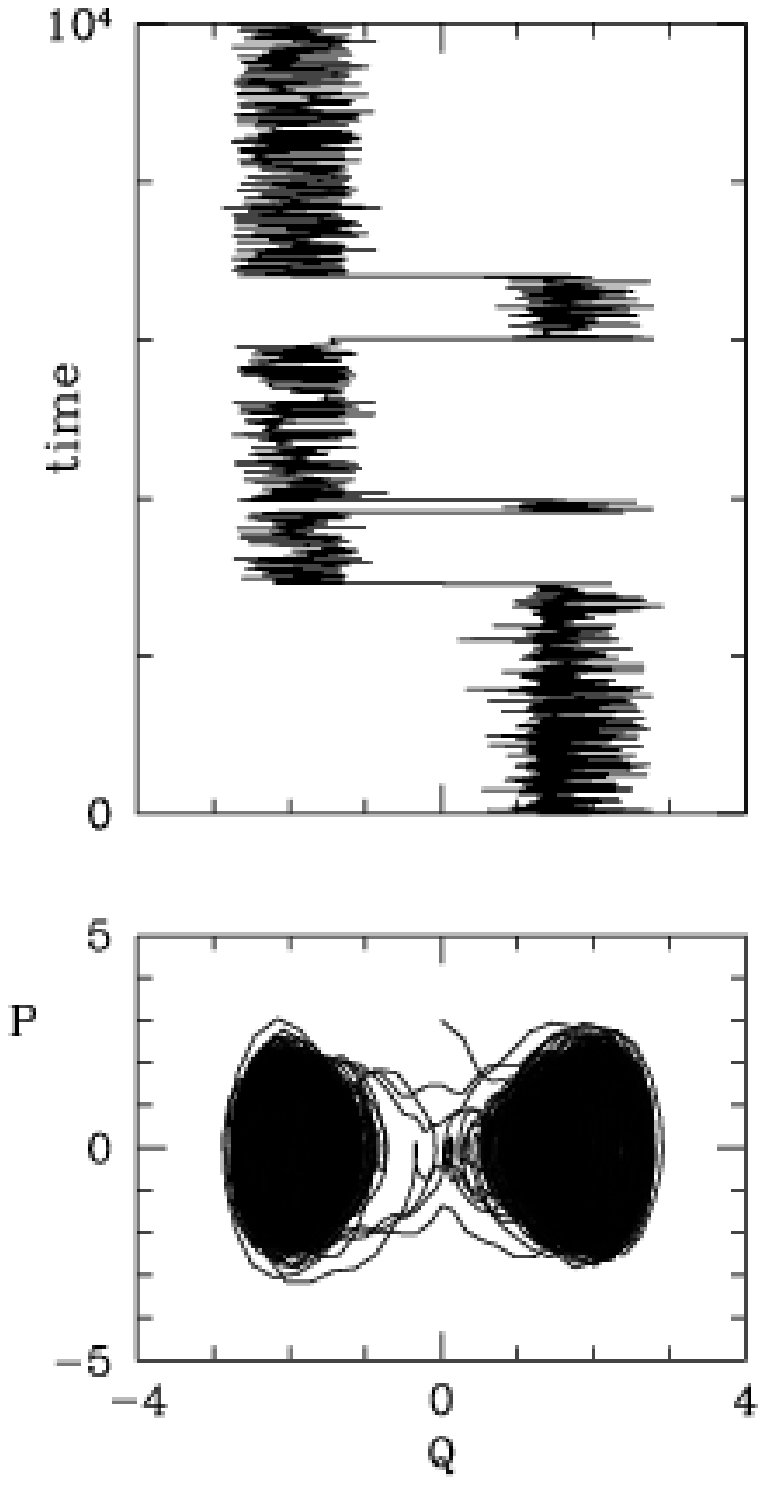}
}
\caption{Typical behavior for the dynamics of the distinguished particle in
contact with a bath with $N=8000$ particles. Most of the time the
distinguished particle is located in one of the two potential wells, but
occasionally it makes a jump between wells.}
\label{fig:intro_QP}
\end{figure}

The systems we consider are exemplified by the following model.  A particle,
called the distinguished particle, with position $Q$ and momentum $P$ sits in
a potential well $V(Q)$.  It is coupled via linear springs to a large number
$N$ of particles comprising a heat bath; see Figure~\ref{fig:sketch}.  The
potential well considered here is a slightly asymmetric double well.  The full
system is an $N+1$ degree of freedom Hamiltonian system.  A detailed
description of the model, including the choice of spring constants, masses,
and initial data for the bath particles is given below and in \cite{FK, KSTT,
Zw}.  The important point is that the dynamics of this simple model (and the
others that we consider in this paper) is typical of many more complex
molecular and stochastic systems in which the state is primarily confined to a
few conformations (here defined by the minima of $V$) with rare switching
events between them.
This is illustrated in Figure~\ref{fig:intro_QP} with a
typical trajectory and time series for the distinguished particle.

\begin{figure}
\centerline{
\includegraphics[width=4.0in]{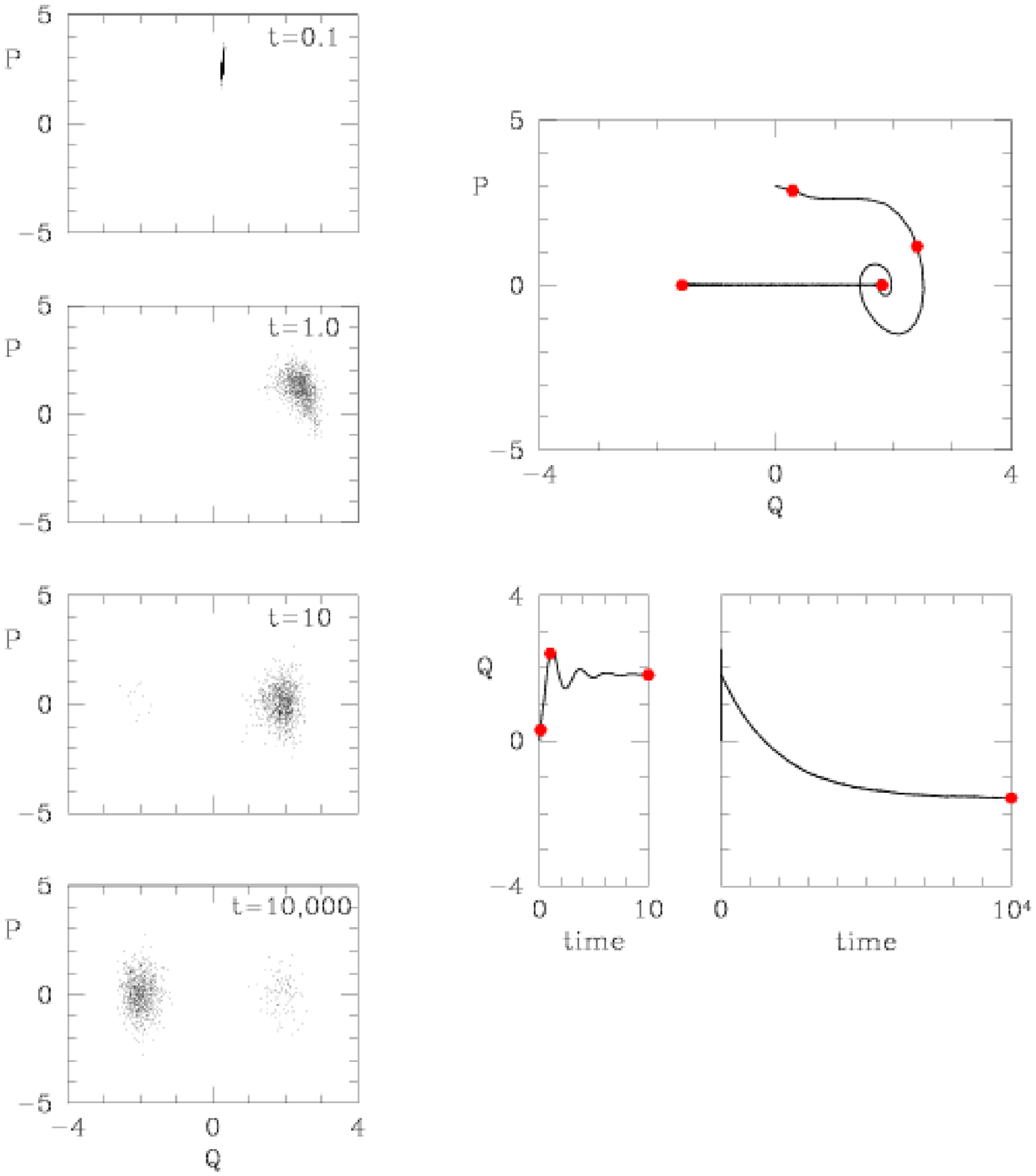}
}
\caption{Evolution of an ensemble of $10^4$ realizations for the model system.
    On the left density plots of position and momentum of the distinguished
    particle are shown at four times as labeled.  For clarity only the first
    $10^3$ realizations are plotted. The initial conditions are $(Q,P) =
    (0,3)$.  The right shows the trajectory and time series for the ensemble
    expectations.  Red points indicate the four times shown on the left.
    There are $N=8000$ particles in the heat bath. }
\label{fig:intro_ensemble}
\end{figure}

Consider now the dynamics of an ensemble of trajectories for the model system.
Figure~\ref{fig:intro_ensemble} illustrates the evolution of an ensemble of
trajectories all with the same initial conditions $(Q,P) = (0,3)$ for the
distinguished particle, but with different initial data for the bath
particles.  (The total initial energy of the bath is approximately the same
for all realizations. See \S\ref{sec:models}.)  Over a time of order $10$ the
initial density evolves to a nearly Gaussian density centered near the bottom
of the right well, where it remains roughly constant for some time.  While not
immediately evident, during the initial $10$ time units, the density is never
very far from Gaussian.  However, it is evident from the left-hand plot that a
small percentage of the realizations are located in the left well at time
$10$.  Over a much longer time scale, $O(10^4)$, the density evolves to the
bimodal equilibrium distribution and is hence far from Gaussian.  Thus the
system exhibits {\em metastability} with near equilibration within one well
dominating over medium time scales, before the system ultimately converges to
an equilibrium distribution which sees both wells.  The time it takes the
system to reach equilibrium clearly depends on the potential barrier height;
the time scale of the intermediate time evolution to the well bottom depends
only on properties of the particular well.

Our aim is to study the behavior of coarse dynamics and coarse bifurcation
methods on problems which exhibit metastable behavior of this type.
In \S 2 we introduce a hierarchy of model problems, all of which exhibit
rare transitions between a small number of states, and which we then use
throughout as illustrations.

\begin{figure}
\centerline{
\includegraphics[width=2.0in]{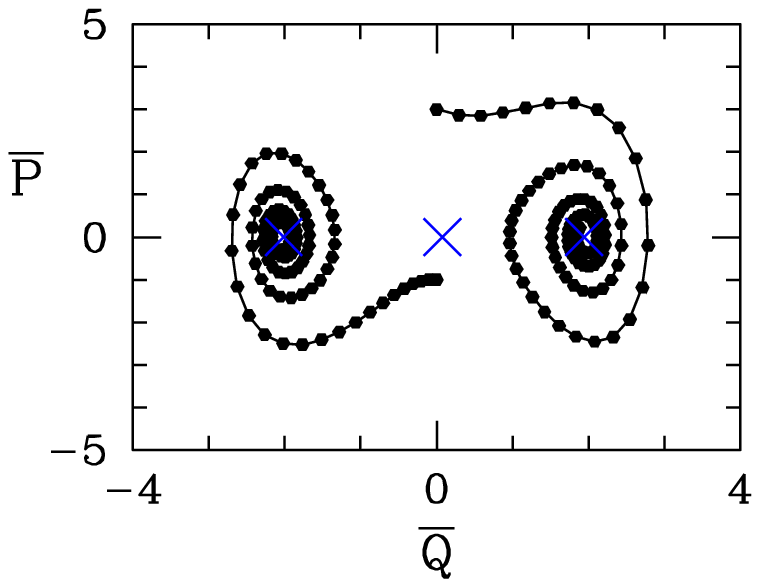}
}
\caption{Dynamics of a moment map for the first-order moments of the position
    $Q$ and momentum $P$ of the distinguished particle. The map has three
    fixed points; two stable foci and one unstable saddle shown as blue
    cross. Two trajectories are shown; one evolving to each of the stable
    fixed points. The right trajectory has initial condition $(\bar Q, \bar P)
    = (0,3)$, corresponding to that in figure~\ref{fig:intro_ensemble}. There
    are $N=8000$ particles in the heat bath. }
\label{fig:intro_moments}
\end{figure}

In \S 3 we define the discrete-time {\em moment map} $\Phi$ for the first $k$
moments of the ensemble of solutions to a time evolving system.  Specifically,
$\Phi$ will be a low-dimensional map for only the low-order moments, defined
in a general setting which applies to both systems of ordinary differential
equations (ODEs), with randomness from initial data, and stochastic
differential equations (SDEs) with randomness from initial data and Brownian
driving noise.  Figure~\ref{fig:intro_moments} illustrates the {\it
first-order moment map} for the heat bath example.  The short-term dynamics of
the map resemble those of the ensemble, but significantly, the moment map has
{\em stable fixed points} corresponding to means of metastable densities
centered in each well.  The second-order moment map additionally captures the
widths of the metastable measures.  The algorithms we study are based on these
maps.

The heart of the paper is \S 4. It is devoted to the study of the moment map,
when applied to a variety of model systems.  We use a combination of exact
solutions for Gaussian problems, approximate solutions for metastable systems
and numerical experiments.  Of central interest is the observation that the
moment maps stabilize the metastable states in the model problems of interest.
The moment map is a nonlinear map, defined from the linear flow of probability
densities for ODEs and SDEs, with the nonlinearity entering through the
process of repeatedly projecting onto the space of moments.  This process of
{\em nonlinearization} creates interesting fixed points which are associated
with metastable behavior, and are amenable to low dimensional bifurcation
analyses; related issues are addressed in \cite{HKRT94}.  \S 5 contains our
concluding remarks.

%

\section{Model Problems}
\label{sec:models}

We consider three example systems in this paper.  Two are SDEs and one is the
ODE heat bath model described in the introduction, and illustrated in
Figure~\ref{fig:sketch}.  A major thrust of this paper is to establish,
through computational experimentation, that the moment map stabilizes
metastable behavior arising from the slow dynamics between potential wells
with large energy barriers; this gives rise to nonlinear phenomena, such as
bifurcations, in the moment map.  Such phenomena can be illustrated in both
SDE models, and in the ODE heat bath model.  Furthermore, in various parameter
limits, the SDEs can be derived as approximations for the heat bath, further
justifying their study.

\subsection{Heat-bath model}
\label{ssec:heat}

This model problem is defined by the Hamiltonian
\begin{equation}
H(Q,P,q,p) = \frac{1}{2M} P^2 + V(Q) +
\sum_{j=1}^N \frac{p_j^2}{2 m_j} + 
\sum_{j=1}^N \frac{k_j}{2} (q_j - Q)^2,
\label{eq:H}
\end{equation}
where $Q,P$ are the position and momentum of a distinguished particle of unit
mass in a potential field $V(\cdot)$. The $q_j$'s and $p_j$'s are the
coordinates and momenta of $N$ particles that are referred to as ``heat bath''
particles. The $j$-th heat bath particle has mass $m_j$ and interacts with the
distinguished particle via a linear spring with stiffness constant $k_j$. If
the distinguished particle were held fixed it would be the anchor point of $N$
independent oscillators with frequencies $\w_j = (k_j/m_j)^{1/2}$.
The numerical experiments are all conducted with mass $M=1.$

Initial conditions for the distinguished particle are: $Q(0)=Q_0$, $P(0)=P_0$.
The initial data for the heat bath particles, $q_j(0)=q_j^0$ and
$p_j(0)=p_j^0$, are randomly drawn from a Gibbs distribution with inverse
temperature $\beta$. The Gibbs measure is conditioned by the (non-random)
initial data $Q_0$ and $P_0$. For fixed $Q,P$ the Hamiltonian \eqref{eq:H} is
quadratic in $q,p$, and hence the corresponding measure is Gaussian. It is
easily verified that
\[
\begin{aligned}
q_j^0 &= Q_0 + (1/\beta k_j)^{1/2} \xi_j \\
p_j^0 &= (m_j/\beta)^{1/2} \eta_j,
\end{aligned}
\]
where $\xi_j,\eta_j\sim\calN(0,1)$ are mutually independent sequences
of i.i.d.\ random variables.

This leaves the specification of the parameters $k_j$ and $m_j$.  For our
purposes the only important property of these parameters is that the
frequencies $\w_j = (k_j/m_j)^{1/2}$ cover an increasingly large range in an
increasingly dense manner as the number of particles $N$ increases.  Hence what
we actually specify is the frequencies.  These are chosen to be random and
uniformly distributed in $[0,N^{1/3}]$,
\[
\w_j = N^{1/3}\nu_j, \quad \nu_{j}\,\,\, i.i.d., \quad
\nu_1 \sim\calU[0,1].
\]
It is important to note that in addition to the initial data, the model
specification itself contains this random element. We shall be careful to
distinguish between the two types of randomness. 

From the frequencies the spring constants and masses are given by:
\begin{equation}
k_j = \frac{2\alpha^2}{\pi(\alpha^2 + \w_j^2)} \, \frac{N^{1/3}}{N},
\qquad m_j = \frac{k_j}{\w_j^2}
\label{eq:kj}
\end{equation}
with $\alpha>0$. See \cite{KSTT} for further details.

The parameters $\alpha$ and $\beta$ are fixed at $\alpha = 100, \beta = 2.$
The potential considered in this paper is
\begin{eqnarray}
\label{eq:pot}
V(Q) = \frac{Q^4}{4} - \frac{\mu Q^2}{2} + \nu Q
\end{eqnarray}
where $\mu$ and $\nu$ are parameters with $\nu$ typically small.

\subsection{Two Dimensional SDE Approximation}
\label{ssec:2DSDE}

For large $N$ and $\alpha$ the distinguished particle $Q$ in the 
heat bath model can be approximated by the SDE
\begin{equation}
\label{eq:SDE2}
M\ddot{Q}+\gamma \dot{Q}+V'(Q)=\sqrt{2\gamma/\beta}\;\dot{W}.
\end{equation} 
A theorem justifying this approximation can be proved using the techniques
of weak convergence, by taking the limit $N \to \infty$ (\cite{KSTT}) and
then $\alpha \to \infty$ (\cite{PS}). In the absence of noise
(the zero temperature limit $\beta \to \infty$) this damped Hamiltonian
system exhibits
decay towards stationary points with zero velocity and positions
at the critical points of $V.$ The presence of noise (finite $\beta$)
then induces transitions
between the minima of $V$, with time-scales determined by
the well-depths relative to the size of noise.

\subsection{One Dimensional SDE Approximation}
\label{ssec:1DSDE}

The stochastic dynamics between potential wells are also present in simple one
dimensional SDEs. A particular instance of such a one dimensional SDE follows
from \eqref{eq:SDE2} for $M \ll 1$.  In the limit $M \to 0$, the solutions of
\eqref{eq:SDE2} converge strongly \cite{Nelson} or weakly \cite{Gardiner} to
solutions of the SDE
\begin{equation}
\label{eq:SDE1}
\gamma \dot{Q}+V'(Q)=\sqrt{2\gamma/\beta}\dot{W}.
\end{equation} 
We will use this problem to illustrate the moment map, and its properties
on systems exhibiting metastable dynamics within potential wells. 

\subsection{Remark}

In most respects the SDE systems derived above are simpler to treat, and
computationally they are far less expensive to simulate than the full heat
bath model.  Therefore, when we later use these models as examples, we will
study the models in the opposite order from what has just been presented.  We
start with the 1D SDE and examine its behavior extensively and then consider
more briefly 2D SDE and the heat bath system of ODEs.

\section{The Moment Map}
\label{sec:map}

The central objects of our study are maps on moments.  The basic ingredients
are an evolution equation (either an SDE or system of ODEs), a space of
low-order moments, and a measure determined uniquely by low-order moments.  We
refer to the latter as {\em lifting} and choice of the lifting operator is an
essential ingredient in the method. We start with the SDE case, then describe
the situation for ODEs.  For both the SDE and the ODEs the flow on probability
densities is linear.  The nonlinearity inherent in the moment map comes from
the relationship between the probability density function and its moments.
After describing the moment map for the SDE and the ODE, the section concludes
with some general remarks.

\subsection{The SDE Case}
\label{ssec:sde}

Let $x \in C([0,\infty),\RR^d)$ solve the following It\^{o} SDE, driven by
Brownian motion $W \in C([0,\infty),\RR^m):$
\begin{equation}
\label{eq:sde}
\frac{dx}{dt}=f(x)+\sigma(x)\frac{dW}{dt}. 
\end{equation}
This includes \eqref{eq:SDE2} and \eqref{eq:SDE1} as special cases.  We will
consider ensembles of solutions of this equation, with ensembles taken over
multiple driving noises and random initial data.  Let $X_j(t)$ be the $j^{th}$
moment of $x(t)$, with expectation taken with respect to both the driving
Brownian motion $W$ and random initial data, the latter being assumed
independent of the Brownian motion.  Denote the first $k$ moments of $x$ by
$X(t)=(X_1(t), \dots, X_k(t))$.

Let $\mu$ be a measure on $\RR^d$ determined by exactly $k$ moments, with
density $\hat\rho(x;X).$ Here $X$ denotes the dependence of the density on the
$k$ moments and we require that the $k$ moments of $\hat \rho$ are exactly
those given by $X.$

We now define the map $\Phi$ on $k$ moments by
\begin{equation}
\label{eq:map}
\Phi(X)=X(\tau)
\end{equation}
where the initial data $X(0)=x_0$ is distributed with density $\hat\rho(x;X)$
and $\tau \le [0,<\infty).$ (Note that if $\tau=0$ then $\Phi$ is the
identity.)  We refer to this as the {\em moment map}.  It depends on three
choices that need to be made when calculating the moment map:

\begin{enumerate}[(i)]
\item the evolution time $\tau$; 
\item the {\it number} of moments we choose to use; 
\item the lifting step: the way choose to distribute the
initial density based on the moments.
\end{enumerate}

We will return to this dependence in more detail below.
The two examples of the measures which will be used
throughout this paper are the Dirac measure and the Gaussian measure, uniquely
determined by the first moment and the first two moments, respectively.

In practice we typically find this map through Monte Carlo
simulation, but it is insightful to describe the definition of
the map through the Fokker-Planck equation for \eqref{eq:sde}. This
linear PDE for the probability densities $\rho(x,t)$ propagated by
\eqref{eq:sde} is 
\begin{align}
\label{eq:fp}
\frac{\partial \rho}{\partial t} = - \nabla \cdot(f\rho) + 
\frac{1}{2} \nabla \cdot {\nabla \cdot (\Sigma\rho)}:=
& {\cal L}^*\rho,\\
\rho(x,0)=&\hat\rho(x;X)
\end{align}
where $\Sigma=\sigma(x)\sigma(x)^T.$ Here ${\cal L}^*$ is the adjoint of the
generator for the process ${\cal L}.$ We denote the solution by
\begin{equation}
\rho(x,t)=e^{{\cal L}^*t}\hat\rho(x;X).
\label{eq:prop}
\end{equation}
From $\rho(x,\tau)$ we can construct $\Phi(X)$ by \eqref{eq:map}.  In general
the moment map is {\em nonlinear} because of the nonlinear dependence of
$\hat\rho(x;X)$ on $X$ and the nonlinearity of the map from $\rho(x,t)$ to
$X(\tau)$ given by \eqref{eq:prop}.  Thus we have constructed a nonlinear map
on $\RR^d$ from the linear flow on $\RR^d-$valued densities.

\begin{figure}
\centerline{ \includegraphics[width=3.5in]{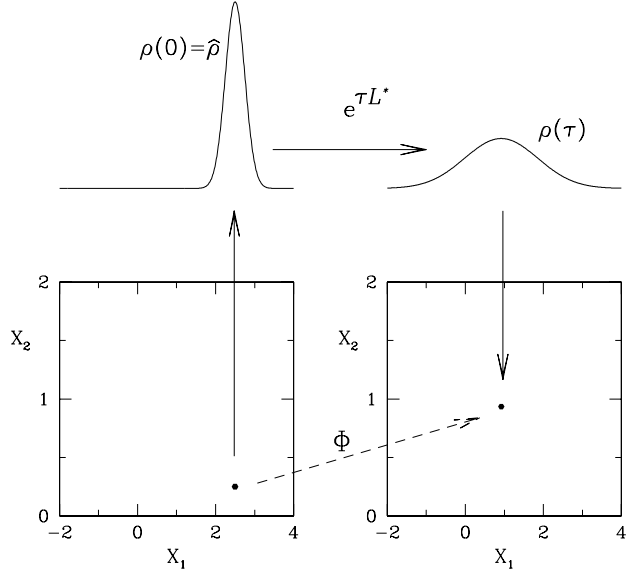} }
\caption{Sketch illustrating the definition of the moment map.  From a point
  $(X_1, X_2)$ in moment space a density $\hat\rho$ is uniquely
  determined. This initial density is evolved by the system dynamics (e.g. the
  Fokker-Planck equation).  From the resulting density $\rho(\tau)$, $\tau$
  time units later, a new point in moment space is determined.  This point is
  defined to be the image of $(X_1, X_2)$ under the moment map.  }
\label{fig:mm}
\end{figure}

It is helpful to consider Figure~\ref{fig:mm} illustrating the definition of
the moment map.  One should view the map as a composition of three steps: (1)
{\it lifting} from a space of moments (subset of $\RR^d$) to a space of
probability densities (captured by our choice of $\hat\rho(\cdot;X)$); (2)
time evolution of the density by the underlying process (a linear flow, given
by the map $e^{{\cal L}^*\tau}$); (3) projection back to moment space by
integrating against the time-evolved measure.

In the case of a single moment, in this paper, we take $\mu$ to be a Dirac
measure at $X$ and then $X(\tau)=\bbE x(\tau)$ can be calculated from
$$\Phi(X)=\int_{\RR^d} xe^{{\cal L}^*\tau}\delta(x-X)dx.$$ In the case of two
moments we have
$$
X(\tau)=\{\bbE x(\tau), 
\bbE [x(\tau)-\bbE x(\tau)][x(\tau)-\bbE x(\tau)]^T \}
$$
and we take $\mu$ to be a Gaussian measure with mean and covariance determined
by these moments.  It is convenient to express the moment map in terms of the
mean $\bar X \in \bbR^d$ and covariance matrix $\Sigma \in \bbR^{d \times
d}$. We obtain the map
\begin{eqnarray*}
\Phi(\bar X,\Sigma)=\left\{
\begin{array}{c}
\Phi_1(\bar X,\Sigma)\\
\Phi_2(\bar X,\Sigma)
\end{array}
\right\}
\end{eqnarray*}
The functions $\Phi_i$ are defined as follows. Let
$$
\hat{\rho}(x;\bar X,\Sigma)=\frac{\exp(-\frac12\|
\Sigma^{-\frac12}(x-\bar X)\|^2)}
{\sqrt ((2\pi)^d\det \Sigma)}.
$$
Then $\Phi_1:\RR^d \times \RR^{d \times d} \to \RR^d$ is given by
$$
\Phi_1(\bar X,\Sigma)=\int_{\RR^d}x\{e^{{\cal L}^* \tau}
\hat{\rho}(x;\bar X,\Sigma)\}dx
$$
and $\Phi_2:\RR^d \times \RR^{d \times d} \to \RR^{d \times d}$
is given by
$$\Phi_2(\bar X,\Sigma)=
\int_{\RR^d}(x-\Phi_1(\bar X,\Sigma))(x-\Phi_1(\bar X,\Sigma))^T
\{e^{{\cal L}^* \tau} \hat{\rho}(x;\bar X,\Sigma)\}dx.
$$ 

There is a connection between particle filters and the moment map, although
the former represent the desired density as a sum of several delta functions,
or Gaussians, not just as one \cite{Doucet}.

\subsection{The ODE Case}
\label{ssec:ode}

The moment map can be defined for deterministic problems of the form
\begin{align}
\label{eq:ode}
\frac{dx}{dt}&=f(x,y),\\
\frac{dy}{dt}&=g(x,y).
\end{align}
Here $x \in \RR^d$ and $y \in \RR^m$ and the randomness is assumed to come
entirely from the initial data.  In systems characterized by a separation of
time scales, it is sometimes the case that one can write an effective reduced
models in terms of a subset of (typically slow) variables.  Under such
appropriate conditions, we might, for example, be interested in finding a map
in terms of the first $k$ moments of $x$ alone.  Thus the measure $\mu$ must
be chosen on $\RR^d \times \RR^m$ so that it is uniquely characterized by $X$,
the first $k$ moments of $x \in \RR^d$.  It is natural to choose $\mu$ to be
an invariant measure for the flow, conditioned by knowledge of the first $k$
moments of $x$; if the flow is Hamiltonian then a Gibbs measure is often
used. We denote the density associated with this measure by $\hat\rho(x,y;X).$
This occurs in the heat bath example considered in the introduction where $x$
represents coordinate and momentum of the distinguished particle, whilst $y$
represents the heat bath coordinates and momenta.

Rather than the Fokker-Planck equation \eqref{eq:fp} we have the Liouville
equation for propagation of probability densities. This is
\begin{align*}
\frac{\partial \rho}{\partial t} = - \nabla_{x} \cdot(f\rho) -
\nabla_{y} \cdot(g\rho):= &{\cal L}^*\rho,\\
\rho(x,y,0)=&\hat\rho(x,y;X)
\end{align*}
and we denote the solution by
$$
\rho(x,y,t)=e^{{\cal L}^*t}\hat\rho(x,y;X).
$$

In the case of a Dirac mass we take
$\hat\rho(x,y;X)=\delta(x-X)\hat\rho(y;X)$ and then the map
$$
\Phi(X)=\int_{\RR^d \times \RR^m} x\{e^{{\cal L}^*\tau}
\delta(x-X)\hat\rho(y;X)\}dxdy.
$$
Here $\hat{\rho}(y;X)$ is chosen so that
$\delta(x-X)\hat\rho(y;X)$ is the density of $\mu$ conditional on $x=X.$ 
Generalization to Gaussian, and higher moment
problems, is also possible.

\subsection{General Remarks}

\begin{itemize}

\item {\em Notation.} We use $X$ to represent a point in the moment space up
to some order $k$, which will be made explicit for each particular example we
consider.  In practice, the coordinates used to describe the moment space are
dictated by the particular problem.  For example, in our case we use the mean
and standard deviation as coordinates when considering $k=2$ and $d=1$.  For
$k=2$ and $d=2$ we use the two means, the two standard deviations, and the
cross correlation as coordinates.

\item {\em Usage.} In discussing moment maps we often do not distinguish
between a point $X$ in moment space and the uniquely determined density
$\hat\rho(\cdot;X)$ based on this point.  That is, we sometime speak of the
moment map as mapping $X^n$ to $X^{n+1}$ and sometime speak of the moment map
as taking mapping density $\hat\rho(\cdot;X^n)$ to density
$\hat\rho(\cdot;X^{n+1})$.

\item {\em Relation to Optimal Prediction}. The map $\Phi$ can be used to
generate an approximate vector field by defining
\begin{equation}
\label{eq:vf}
F(X)=\frac{\Phi(X)-X}{\tau}.
\end{equation}
For the ODE case \eqref{eq:ode} and $\mu$ a Dirac at $X$, the limit $\tau \to
0$ coincides with the vector field found by the method of optimal prediction
\cite{CHK00}; this is demonstrated in \cite{GKS03}. In general the method of
optimal prediction leads to errors which grow linearly in time $T$
\cite{hal99}. The approach we study here attempts to
overcome this error growth by closing the system with a larger number of
moments.

\item {\em Previous Work} For simple problems in chemical kinetics, which are
modeled by birth-death processes, the equation for the first moment is a
closed ODE, in the limit of a large number of independent particles, and the
moment map studied here then works well in the Dirac mass case
\cite{KMC1,KMC2,KMC3}.  For more complex problems, such as lattice Boltzmann,
a closed effective PDE may sometimes be found, using first and second moments
and, again, the moment map works well in this case
\cite{PNAS,GKT,Bubbles,Radek}.  In this paper we study examples where no
closure is proven to exist, and demonstrate the properties of the moment
map. In particular we study the relevance of fixed points of this map to the
identification of metastable states. Although no rigorous analysis is
presented, the numerical studies show that the moment map has some merit as a
method for elucidating long-term dynamics of large systems, through low
dimensional dynamical systems studies.

\item {\em Lifting} Initialization of the detailed simulation consistently
with coarse-grained observables is the {\it lifting} step in equation-free
computation \cite{PNAS,GKT,Manifesto,Graham}.  This step is obviously not
unique, as there exist many ways of initializing a distribution conditioned on
a few of its lower order moments.  Our choice of a particular measure
depending only on the lower order moments allows for a systematical
initialization of the fine scale dynamics, a concept that goes back to
Ehrenfest \cite{Ehrenfest,Gorban}, and is an important component of our
computational approach.  In the case of systems with metastability, different
effective dynamics will be deduced (different closures will be obtained)
depending (a) on the time scale of the observation (the time horizon of the
simulation $\tau$ with the fine scale solver) and (b) on the nature of the
lifting from the moments (the choice of $\rho$).  Over very short times, and
initializing with a Dirac delta function, the simulation will effectively
sample the local gradient of the well; over medium times, and initializing
{\it within one well}, one will observe equilibration within this well; and
over very long times (no matter what the initialization) one will observe the
approach to the equilibrium density.  If we want to study the system over
medium time scales, it is obviously important to use a time $\tau$ in the
construction of the moment map that is short compared to the escape time
between wells, but long enough for the noise (dynamics) to allow the sampling
of the features of the well bottom. On a longer time-scale it is necessary to
incorporate the transition time between wells as is done, for example, in the
method of conditional averaging \cite{Schuette}.

\item {\em Computational Savings}.
The moment map can lead to computational savings in two primary ways:

\begin{enumerate}[(i)]
\item the map $\Phi$ can be used in finite dimensional bifurcation and
continuation studies;
\item the estimated vector field $F$ can be used to advance the moments over
several multiples of the time-step $\tau$.
\end{enumerate}

In case (i) savings can arise from using accelerated methods, such as Newton
iteration in a continuation environment, to find fixed points. In case (ii)
savings arise by considering maps of the type
$$X^{n+1}=X^n+l\{\Phi(X^n)-X^n\}$$ to advance the moments through $l\tau$ time
units, using solution of the full problem \eqref{eq:sde} or \eqref{eq:ode}
only over $\tau$ time units. (If $l=1$ this simply reduces to the moment map).
The above formula constitutes a {\it projective forward Euler} explicit coarse
integrator.  Much more sophisticated integrators, including multistep and
implicit ones can also be used; a rigorous analysis of savings in case (ii)
has only recently been initiated \cite{KevGear1,KevGear2,KevGear3}.  These
ideas have been applied to a wide range of problems, both deterministic and
stochastic, see
e.g.~\cite{GEAROLD,GKT,KMC1,KMC2,KMC3,Bubbles,Graham,Radek,Mikko,Dima,HUMMER1,SIMA,LEVIN}
and other references in \cite{Manifesto,ShortManifesto}.  An approach related
to case (ii), which can be fully optimized for variance reduction and so forth
when explicit time-scale separation occurs between the $x$ and $y$ dynamics in
\eqref{eq:vf}, is outlined in \cite{vde03}; for a rigorous analysis see
\cite{ELvde04}.

\end{itemize}

\section{Examples}
\label{sec:anex}

Here we explore a series of examples of moment maps. We start with examples
based on the OU process for which explicit representation is possible. In
these cases the moment maps are linear.  We then proceed to the more
interesting nonlinear maps arising from systems with double-well
potentials. 

\subsection{The OU Process} 

Consider the OU process 
\begin{equation}
\label{eq:ou}
\frac{dx}{dt}=-\alpha x+\sqrt \lambda \frac{dW}{dt}.
\end{equation}
This is essentially the simplest example of \eqref{eq:sde} and corresponds
to \eqref{eq:SDE1} in the case of a quadratic potential.

The exact solution of this process is
\begin{equation}
\label{eq:ousol}
x(t)=e^{-\alpha t}x(0)+\sqrt \lambda \int_0^t e^{-\alpha(t-s)}dW(s).
\end{equation}

\begin{figure}
\centerline{
\includegraphics[width=4in]{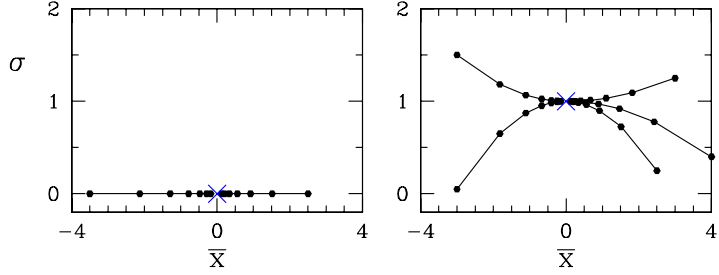}
}
\caption{ Phase portraits for first-order (left) and second-order (right)
moment maps for the OU process.  
Trajectories for these linear maps are shown
starting from several different initial conditions. Parameters are $\alpha=1$,
$\lambda=2$, and $\tau=0.5$. }
\label{fig:OU_mm}
\end{figure}

For first-order moment map on $\bar{x}$, we take initial data with Dirac
measure with density ${\hat \rho}(x;\bar{x}) = \delta(x-\bar{x})$.  The map on
the first moment $\bar{x}$ is explicitly
$$
\Phi(\bar{x})= \E x(\tau) = \int x\{e^{{\cal L}^*\tau}\delta(x-\bar{x})\}dx.
$$
From \eqref{eq:ousol} we have
\begin{equation}
\label{eq:oumap1}
\Phi(\bar{x})=e^{-\alpha \tau}\bar{x}.
\end{equation}
The map $\Phi$ is linear and has a unique, globally attracting fixed point at
$\bar{x}=0$.  Figure \ref{fig:OU_mm} shows a phase portrait for this
simple map.  

Before discussing this we consider the second-order moment map with mean
$\bar{x}$ and standard deviation $\sigma$ as co-ordinates.
In this case the initial data has density
$$
{\hat \rho}(x;\bar{x}, \sigma) = 
\frac{1}{\sqrt{2\pi}\sigma} e^{-\frac{(x-\bar{x})^2}{2 \sigma^2}}.
$$
From \eqref{eq:ousol} we then have that the moment map on $(\bar{x},\sigma)$
is given by 
\begin{align}
\label{eq:oumap2}
\Phi_1(\bar{x},\sigma)=&e^{-\alpha \tau}\bar{x}\\
\Phi_2(\bar{x},\sigma)=&\{e^{-2\alpha \tau}\sigma^2+\frac{\lambda}{2\alpha}[1-
e^{-2\alpha \tau}]\}^{\frac12}.
\end{align}
While not a linear map on $(\bar{x},\sigma)$, this is linear on
$(\bar{x},\sigma^2)$.  Figure~\ref{fig:OU_mm} shows a phase portrait for map
\eqref{eq:oumap2}. The map has the unique fixed point $(\bar{x},\sigma)=
(0,\sqrt\{\frac{\lambda}{2\alpha}\})$ which is globally attracting.

The solution to \eqref{eq:ousol} is Gaussian if $x_0$ is Gaussian.  Hence the
second-order moment map \eqref{eq:oumap2} gives exact time $\tau$ samples of
the distribution of the SDE.  This unique fixed point of the map characterizes
the unique invariant (Gaussian) measure of \eqref{eq:ou}.  In contrast, the
first-order moment map, with $\mu$ a Dirac, can only approximate the solution.
Although this map does not quantitatively represent the solution, since it
contains no information about the width of the measure, it captures the
correct dynamics of the first-order moment and shows that probability mass
initially far from the origin will be transported inward.

\subsection{1D SDE: The Double Well Potential}
\label{sec:double}

We now present a detailed study of moment maps for the one-dimensional SDE
\eqref{eq:SDE1} with double-well potential. In this section we focus on the
dynamics of these maps using numerical simulations. In the next section we
analyze the maps, in particular the nonlinearity of the maps, in more detail.

After suitable scaling SDE \eqref{eq:SDE1} can be rewritten in the notation
of \eqref{eq:sde} as
\begin{equation}
\label{eq:SDE1b}
\dot{x}= -V'(x) + \dot{W}.
\end{equation} 
We consider the double-well potential 
\begin{equation}
\label{eq:SDE1bb}
V'(x) = x(x^2-\mu) + \nu. 
\end{equation} 
For $\nu=0$ the potential is symmetric about zero and for $\nu$ small this
symmetry is weakly broken.  The potential has two local minima for $|\nu| < 2
(\mu/3)^{3/2}$ and one minimum otherwise.  

The first-order (Dirac-based measure) and second-order (Gaussian-based
measure) moment maps for this equation are {\it nonlinear}.  In particular
these maps have multiple stable fixed points which undergo bifurcations as the
potential ($\mu$ or $\nu$) is varied.  We are interested in these fixed points
and their stable and unstable manifolds as a function of $\mu$ for $\nu$
fixed. We shall consider two cases, the slightly asymmetric case $\nu=0.3$ and
the symmetric case $\nu=0$.  We resort solely to numerical studies of the
moment maps throughout this section.  In brief, we use Monte Carlo simulations
to evolve densities forward over time interval $\tau$, as in the evolution
from $\hat \rho$ to $\rho$ in figure~\ref{fig:mm}. This numerically determines
the moment maps. By employing additional techniques we can compute steady
states and bifurcations.  The effective fixed point, bifurcation and
continuation calculations require estimates of the Jacobian of the moment map
(or its action). This is achieved by using nearby initializations of the
moment map (see \cite{KMC1,KMC2} as well as \cite{PNAS,Manifesto,Kelley} and
the monograph \cite{Kelleybook} for matrix-free implementations of
Newton-GMRES).

\begin{figure}
\centerline{
\includegraphics[width=4in]{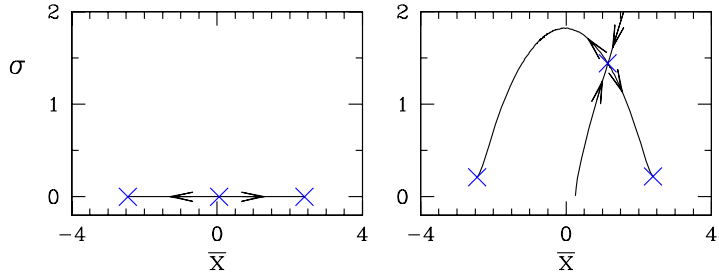}
}
\caption{Phase portraits for first-order (left) and second-order (right)
moment maps for the one-variable SDE in the case of a slightly asymmetric
potential.  Fixed points are indicated by crosses. The stable (for
second-order map) and unstable manifolds of the middle fixed point are
shown. Note, the stable manifold of the saddle in the second-order map does
not pass through the middle fixed point of the first-order map.  Parameters
are $\mu=6$, $\nu=0.3$, $\tau=0.1$.}
\label{fig:SDE1_pp}
\end{figure}

We first consider the moment maps for the slightly asymmetric potential with
representative values of $\mu$ and map time $\tau$.  Figure~\ref{fig:SDE1_pp}
shows phase portraits for both first-order and second-order moment maps.  Each
map has three fixed points. These are shown together with unstable, and for
the second-order map also stable, manifolds of the unstable fixed point.

\begin{figure}
\centerline{
\includegraphics[width=4in]{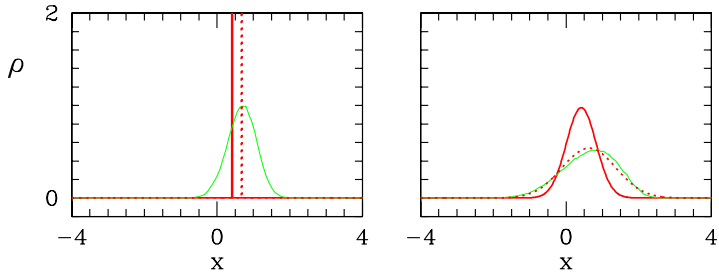}
}
\caption{One iteration of the first and second order maps whose phase
  portraits are shown in figure~\ref{fig:SDE1_pp}.  The bold red curve shows
  $\rho(0) = \hat\rho(x;X^n)$, the thin green curve shows $\rho(\tau)$,
  and the dash bold red curve shows $\hat\rho(x;X^{n+1})$.  (See
  figure~\ref{fig:mm}.)}
\label{fig:SDE1_phi}
\end{figure}

Figure~\ref{fig:SDE1_phi} shows one iteration of both the first- and
second-order maps.  Consider the first-order map. The initial density $\rho(0)
= \hat\rho(x;\bar x)$ is a Dirac delta at $\bar x$, here a point slightly to
the right of the unstable fixed point near zero. After time $\tau=0.1$ the
mean of the density has moved to the right and hence the map takes $\bar x$ to
the right in this case.  The density spreads considerably but over $\tau=0.1$
it remains nearly symmetric (the mean is indistinguishable from the maximum).
The initial density for the next iteration is a Dirac delta displaced to the
right.

For the second-order map the initial density $\rho(0) = \hat\rho(x;\bar x)$ is
a Gaussian centered at $\bar x$ with width $\sigma$.  Here $(\sigma, \bar x) =
(0.418, 0.409)$ corresponds to a point on the lower branch of the stable
manifold of the saddle fixed point. After $\tau=0.1$ time units the density
has spread and the mean has moved slightly to the right.  The map thus
corresponds to substantial increase in $\sigma$ and small increase in $\bar
x$. The density $\rho(\tau)$ is slightly non-Gaussian as can be seen in
comparison with the initial (Gaussian) density for the next iteration.

\begin{figure}
\centerline{
\includegraphics[width=4in]{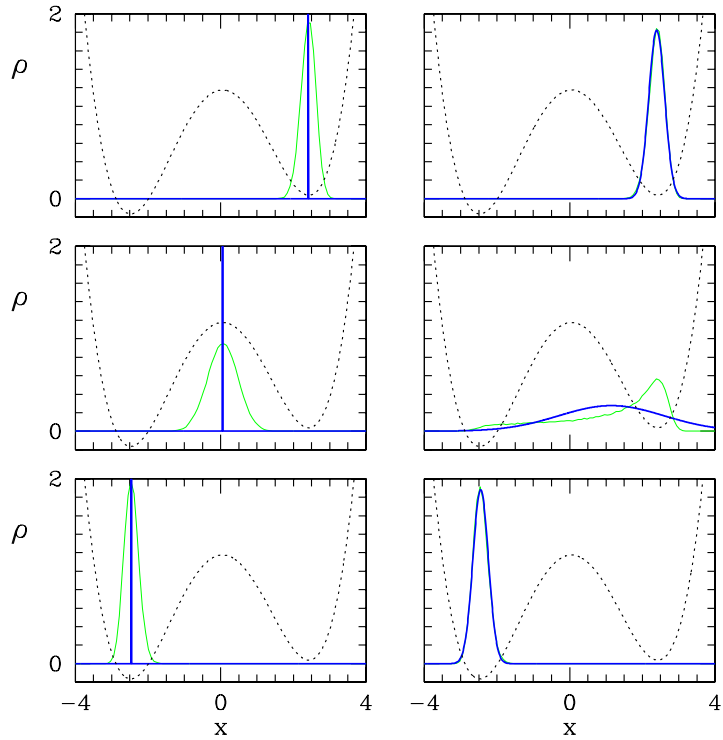}
}
\caption{Densities $\hat\rho(\cdot;X)$, shown with bold blue curves,
corresponding to fixed points for first-order (left) and second-order (right)
moment maps shown in Figure \ref{fig:SDE1_pp}.  Thin green curve show the
evolved density $\rho(\tau)$ for each case. (For the two stable fixed point of
the second-order moment map the evolved density $\rho(\tau)$ is
indistinguishable from $\hat\rho(\cdot;X)$.)  The potential $V(x)$ is shown as
a dashed curve.  }
\label{fig:SDE1_fixed}
\end{figure}

Figure~\ref{fig:SDE1_fixed} shows all fixed points in
Figure~\ref{fig:SDE1_pp}. In each case we plot the density $\hat\rho(\cdot;X)$
corresponding to each fixed point in moment space as well as the density
$\rho(\tau)$.  The stable fixed points of the second-order map are the
metastable measures centered in each well. Specifically, the Gaussian measures
$\hat \rho(x;\bar x, \sigma)$ corresponding to the stable fixed points are
indistinguishable from the evolved densities $\rho(\tau)$. As we shall show
this is independent of $\tau$ over a very large range of $\tau$. Intuitively
this is because on any time scale, up to the very long time scale needed to
reach the equilibrium distribution, these Gaussian measures are approximately
invariant.  The stable fixed points of the first-order map are not invariant
measures.  However, starting from initial condition $\hat \rho(x;\bar x) =
\delta(\bar x)$ where $\bar x$ is indistinguishable from the potential
minimum, the density simply fills out (symmetrically), the (locally quadratic)
well bottom. Note that $\tau=0.1$ is close to, but not quite, the time
necessary to reach the metastable density starting from the Dirac measure.

The unstable fixed points for the maps are understood as follows. For the
first-order map, the unstable fixed point is at the local maximum of the
potential. Starting from a Dirac delta $\hat \rho(x;\bar x) = \delta(\bar x)$,
the density spreads symmetrically, since the maximum is locally quadratic and
hence symmetric.  Hence after time $\tau=0.1$ the mean is still at local
maximum of the potential.  Only for times $\tau$ long enough for the density
to fill the two wells, and hence have a mean different from the local maximum,
would the fixed point be different from the local maximum.  For such times the
fixed point would approximately be the mean of the equilibrium density.  For
the second-order map, the saddle fixed point corresponds to a Gaussian $\hat
\rho(x;\bar x,\sigma)$ which is quite broad.  The evolved density $\rho(\tau)$
is far from Gaussian; it simply has the property that its first two moment
agree with those of the initial Gaussian.  The saddle fixed point is
quantitatively sensitive to the map time $\tau$ (see below).  Qualitatively,
however, for any value of $\tau$ the saddle fixed points are broad Gaussians.

The {\em stable} fixed points of both maps correspond to metastable states
(measures) of the underlying process.  The metastable states are very nearly
Gaussian measures, because the wells are locally quadratic, and they are thus
well captured by the low dimensional moment maps: the densities corresponding
to the stable fixed points of the second-order map are virtually
indistinguishable from the metastable states.  The fixed points of the
first-order map capture the means of the metastable distributions.  The stable
fixed points are insensitive to the value of $\tau$ over a large range of
$\tau$.  (See Figure \ref{fig:SDE1_bd_tdepend} below).  This lack of
sensitivity to $\tau$ suggests that these fixed points are meaningful
characteristics of the observed dynamics over a range of observation time
scales.  The unstable (saddle) fixed point for the first-order moment map is
also insensitive to the value of $\tau$ but the unstable fixed point for the
second-order map is sensitive to $\tau$.  This suggests that the fixed points
of the first-order moment map provides a useful description of the dynamics
(for relatively short times) close to both the saddle and the well bottoms;
the second-order map fixed points provide a meaningful description of the
effective dynamics close to the bottoms of the two wells, {\it but not in the
neighborhood of the saddle}; this is essentially because a reduced equation in
terms of the second-order moments does not appear to successfully close in the
neighborhood of the saddle.

\begin{figure}
\centerline{
\includegraphics[width=4in]{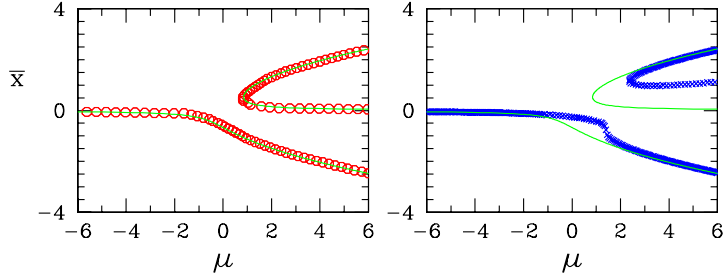}
}
\caption{Bifurcation diagram for first-order (left) and second-order (right)
moment maps for the one-variable SDE in the case of a slightly asymmetric
potential.  Lines show local minima of the potential.  }
\label{fig:SDE1_bd}
\end{figure}

\begin{figure}
\centerline{
\includegraphics[width=4in]{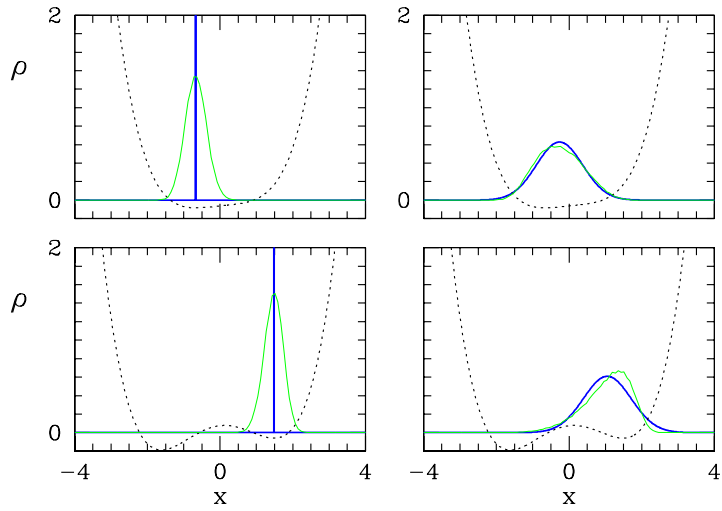}
}
\caption{Densities, shown with bold blue curves, corresponding to fixed points
for first-order (left) and second-order (right) moment maps at two values of
$\mu$ where the non-quadratic aspect of the potential is apparent.  Thin green
curve show the evolved density $\rho(\tau)$ for each case.  The top case is
$\mu=0$. The bottom is $\mu=2.4$, very near the saddle-node bifurcation for
the second-order map in Figure~\ref{fig:SDE1_bd}.  The potential $V(x)$ is
shown as a dashed curve.  }
\label{fig:SDE1_other_fixed}
\end{figure}

We now turn to the behavior of the fixed points as a function of
well-depth. Figure~\ref{fig:SDE1_bd} shows bifurcation diagrams for each
moment map as a function of $\mu$, including $\mu$ for which the potential has
a single well (basically $\mu<0$).  Local extrema of the potential are shown
for comparison.  The right-most end of each bifurcation diagram ($\mu=6$)
corresponds to the phase portraits just considered.  The fixed points for the
first-order map follow the extrema of the potential closely for all $\mu$,
including near the saddle node bifurcation.  The second-order fixed points do
not.  In Figure \ref{fig:SDE1_other_fixed} we show fixed points in cases where
the potential is far from locally quadratic.  One case is $\mu=0$ and the
other is $\mu=2.4$, very near the value corresponding to the saddle-node
bifurcation of the second-order map.  In both cases the first-order fixed
point is at the potential minimum while the mean of the second-order fixed
point is noticeably different from the minimum as can also be seen in
Figure~\ref{fig:SDE1_bd}.

\begin{figure}
\centerline{
\includegraphics[width=4in]{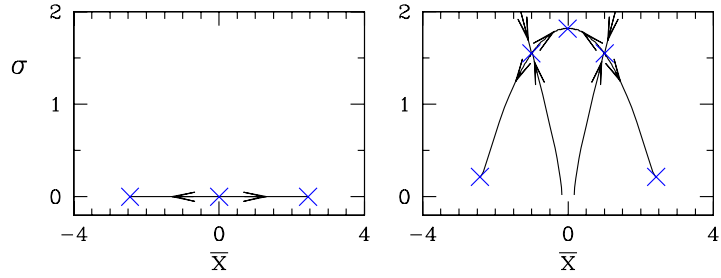}
}
\caption{Phase portrait for first-order (left) and second-order (right) moment
maps for the one-variable SDE in the case of a symmetric potential.  Fixed
points are indicated by crosses. The stable (for second-order map) and
unstable manifolds of the saddle fixed point are shown. }
\label{fig:SDE1_pp_sym}
\end{figure}

We show in figure \ref{fig:SDE1_bd_tdepend} how steady states for the
second-order moment map are affected by the choice of the map time $\tau$.  To
understand what the figure shows, it is helpful to first consider the fixed
points for $\mu<0$.  Neither $\bar x$ nor $\sigma$ vary significantly with
$\tau$ and $\sigma$ is not large.  While not as easy to see, the stable fixed
points for $\mu \gtrsim 2$, behave similarly.  The stable fixed points
correspond to the uppermost and lowermost branches in the $\bar x$-plot and
the bottom branch in the $\sigma$-plot. Both stable fixed points have
approximately the same, relatively small, value of $\sigma$.  The
intermediate, generally unstable, fixed points have larger values of $\sigma$;
these vary quantitatively and qualitatively with $\tau$.  The number of
unstable fixed points can change with $\tau$, e.g.\ at $\mu=6$, $\tau=0.1$
there is one saddle fixed point between the two stable fixed points (figure
\ref{fig:SDE1_pp}) whereas at $\mu=6$, $\tau=1.0$ there are three fixed points
between the two stable points (similar to figure~\ref{fig:SDE1_pp_sym}).  The
conclusion is that the stable fixed points of the second-order moment map
which correspond to stable or metastable measures of the SDE are largely
insensitive to the time $\tau$ used to define the map.  The other fixed points
typically correspond to broad distributions and depend quantitatively on
$\tau$.  This dependence of the fixed points of the moment map on $\tau$
suggests that they are not useful features of the effective dynamics -- which
in turn suggests that the effective behavior {\it does not usefully close} at
the second moment level in the neighborhood of these fixed points.

\begin{figure}
\centerline{
\includegraphics[width=2.5in]{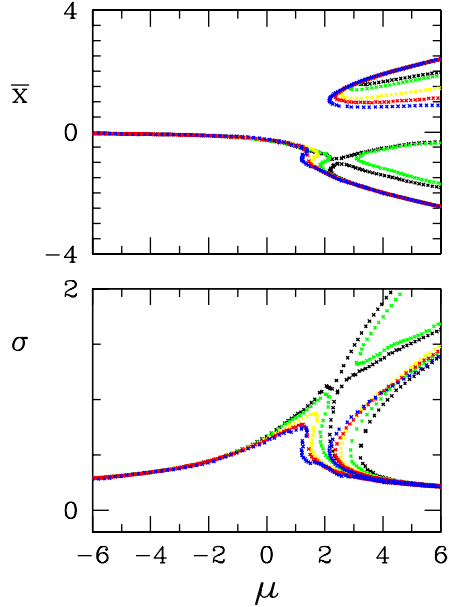}
}
\caption{Bifurcation diagram for second-order moment map for the one-variable
SDE for a variety of values of $\tau$: $\tau=0.05$ blue, $\tau=0.1$ red,
$\tau=0.2$ yellow, $\tau=0.5$ green, $\tau=1.0$ black.  }
\label{fig:SDE1_bd_tdepend}
\end{figure}

\begin{figure}
\centerline{
\includegraphics[width=4in]{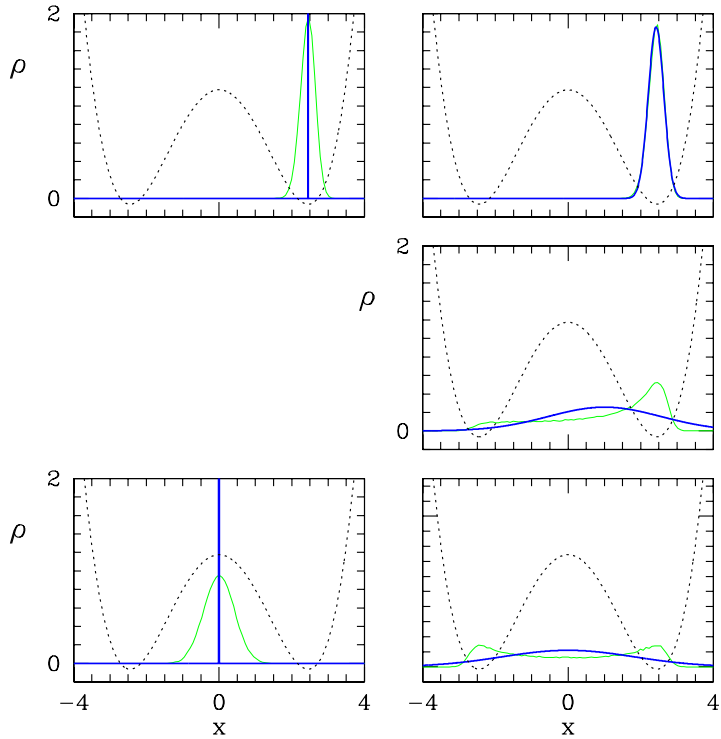}
}
\caption{Fixed points for first-order (left) and second-order (right) moment
maps shown in figure \ref{fig:SDE1_pp_sym}.  Same conventions as in
figure~\ref{fig:SDE1_fixed}.  Top plots show the right stable fixed point. The
middle plot shows the right saddle for the second-order map. The bottom plot
shows the middle fixed point (saddle for the first-order map and stable fixed
point for the second-order map).  The other points
Figure~\ref{fig:SDE1_pp_sym} are obtained by symmetry.  }
\label{fig:SDE1_fixed_sym}
\end{figure}

\begin{figure}
\centerline{
\includegraphics[width=4in]{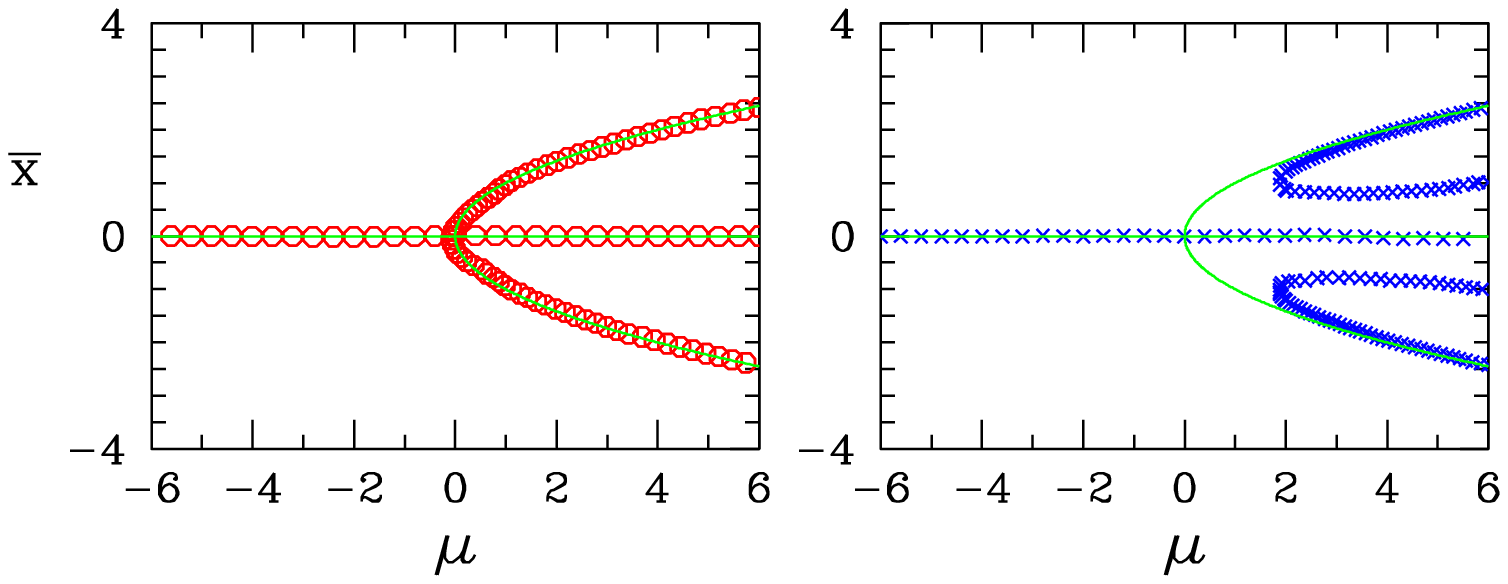}
}
\caption{Bifurcation diagram for first-order (left) and second-order (right)
moment maps for the case of the symmetric potential.  Lines show local minima
of the potential.  }
\label{fig:SDE1_bd_sym}
\end{figure}

Finally we consider the moment maps for the symmetric double-well potential.
Figure \ref{fig:SDE1_pp_sym} shows phase portraits, similar to Figure
\ref{fig:SDE1_pp} for the asymmetric case, while Figure
\ref{fig:SDE1_fixed_sym} fixed points, similar to Figure \ref{fig:SDE1_fixed}
for the asymmetric case.  Figure \ref{fig:SDE1_bd_sym} shows a bifurcation
diagram as a function of $\mu$. An important observation from the data
presented in the symmetric case is that the moment map produces stable
equilibria which are far from metastable states. The Gaussian measures
corresponding to these fixed points are very broad.

\subsection{Metastability and the Double Well Potential}
\label{sec:meta}

We now present some analysis of the nonlinear map.  Consider the SDE
\eqref{eq:SDE1b}.  The adjoint of the generator for this process is
$$
{\cal L}^*\phi(x)=
\frac{d}{dx}\{V'(x)\phi(x)\}+\frac{1}{2}\frac{d^2\phi}
{dx^2}(x).
$$
The equation has a unique invariant density $\rho_{\infty}$, in the null-space
of ${\cal L}^*$, and given explicitly by
\begin{equation}
\rho_{\infty}(x)={\cal Z}^{-1}\exp\{-2 V(x)\},\; {\cal Z}
=\int_{\bbR}\exp\{-2 V(x)\}dx.
\label{eq:imform}
\end{equation}
The operator $\ri^{-\frac12}{\cal L}^*\ri^{\frac12}$ is self-adjoint in the
space $L^2(\bbR)$ (see Proposition 2.2, \cite{Huis}). 
We let $\la \cdot, \cdot \ra$ denote the weighted $L^2(\bbR)$ inner product
$$
\la \theta, \psi \ra=\int_{\bbR}\frac{\theta(x)\psi(x)}{\ri(x)}dx
$$
and we write the eigenvalue problem
$$
{\cal L}^*\phi_j(x)=\lambda_j\phi_j(x)
$$
with eigenvalues ordered so that
$$
0=\lambda_0 \ge \lambda_1 \ge \lambda_2 \ge \dots\;\;.
$$
We may choose the normalization 
$$
\la \phi_j, \phi_k \ra=\delta_{ij}.
$$
Since $\phi_0(x)=\ri(x),$ we have
$$
\int_{\bbR}\phi_0(x)dx=1, \quad \int_{\bbR} \phi_j(x)dx=0,\;j \ge 1.
$$

Now, the solution $\rho(x,t)$ of the Fokker-Planck equation can be expanded
in the eigenfunctions $\phi_{j}$ as
$$
\rho(x,\tau)=\sum_{j=0}^{\infty}a_j(t)\phi_{j}(x),
$$
where
$$
a_j(t) = a_j(0)e^{\lambda_j t}.
$$
Notice that $a_0(0)=0$ in all cases, because
$\rho(x,0)$ is a probability density function.

Assume $V(x)$ in \eqref{eq:SDE1b} is a double well potential with deep wells
relative to the noise.  Then the analysis and numerical evidence
in \cite{Angel, Huis} suggests that it is reasonable to assume that 
$$\epsilon:=\lambda_1/\lambda_2 \ll 1.$$
Thus there exists a spectral gap and, noting that $\lambda_0=0$,
this suggests that after times $\tau$ of order 
$-1/{ \lambda_1}$ the density $\rho(x,t)$ can be well-approximated by 
only the first two eigenfunctions $\phi_0$ and $\phi_1$.

Assuming that $V'' \ne 0$ at the two well bottoms, then $\rho_{\infty}(x)$ is
well-approximated as the weighted sum of two Gaussians $g_{\pm}(x)$; this may
be verified from \eqref{eq:imform}.  We may assume that the Gaussians are
normalized to be probability densities, and then define their means and
standard deviations by
$$\int_{\bbR} g_{\pm}(x)dx=1, \int_{\bbR} xg_{\pm}(x)dx=\bg_{\pm},
\int_{\bbR}[x-\bg_{\pm}]^2 g_{\pm}(x)dx=\bs_{\pm}^2.$$
We then write
\begin{equation}
\label{eq:phi_0}
\phi_0(x)=\rho_{\infty}(x) \approx \alpha g_{+}(x)+(1-\alpha)g_{-}(x),
\end{equation}
where $\alpha$ determines the relative weight of the two Gaussian
contributions. Furthermore, by orthonormality, it may be shown that
\begin{equation}
\label{eq:phi_1}
\phi_{1}(x) \approx \sqrt\{\alpha(1-\alpha)\}[g_{+}(x)-g_{-}(x)].
\end{equation}

We can now use the approximations \eqref{eq:phi_0}, \eqref{eq:phi_1} to
understand the nonlinearity in the moment map for the case of the double well
potential. We focus on the second-order map.  In this case $\rho(x,0) = {\hat
\rho}(x;\bar{x}, \sigma)$ will be Gaussian.  We choose $\tau=-T/\lambda_{1},$
for some $T$ of order $1$ and set
$$\beta=a_1(0)e^{-T}\sqrt\{\alpha(1-\alpha)\}.$$ Then, due to the spectral
gap in the eigenvalues, we have for $\epsilon \ll 1$,
\begin{eqnarray}
\label{eq:approxm}
\begin{array}{ccc}
\rho(x,\tau) &\approx &  \phi_0(x)+a_1(0) e^{-T}\phi_{1}(x)\\
&=& (\alpha+\beta)g_{+}(x)+(1-\alpha-\beta)g_{-}(x)\\
&=& \gamma g_{+}(x)+(1-\gamma) g_{-}(x)
\end{array}
\end{eqnarray}
for $\gamma=\alpha+\beta.$ 

The key to the dynamics of the moment map is manifest in the formula
\eqref{eq:approxm}. Given any Gaussian with mean and standard deviation
$(\bx_0,\s_0)$ we project the density 
$\rho(x,0) = {\hat \rho}(x;\bar{x}_0, \sigma_0)$ onto the basis
$\{\phi_j(x)\}_{j=1}^{\infty}.$ Then the evolution shows that $\rho(x,\tau)$
is given by \eqref{eq:approxm} and
\begin{align}
\bx_1=&\gamma \bg_++(1-\gamma)\bg_-\\
\s_1^2=&\beta\bs_+^2+(1-\beta)\bs_-^2+(1-\beta)\beta (\bg_+-\bg_-)^2.
\end{align}
Since $\gamma$ depends nonlinearly on $(\bx_0,\s_0)$, through $a_1(0)$, we
have a nonlinear map $(\bar{x}_0, \sigma_{0}) \to (\bar{x}_{1},\sigma_{1}).$
In principle this map can be computed explicitly, though this requires knowing
the eigenfunctions $\phi_1$ and $\phi_0$ accurately.

This analysis can be applied to our computational studies of the double well
potential to gain further insight into the moment map.  We employ the
symmetric double well potential considered at the end of \S\ref{sec:double}.
In this case the eigenfunctions $\phi_1$ and $\phi_0$ are accurately
approximated by \eqref{eq:phi_0} and \eqref{eq:phi_1} with $\alpha=1/2$.  We
use $\tau=1$ in the computations which follow. (We shall see that $\lambda_2
\simeq -5$ so that $\tau \lambda_2 \simeq -5$.) 

\begin{figure}
\centerline{
\includegraphics[width=4in]{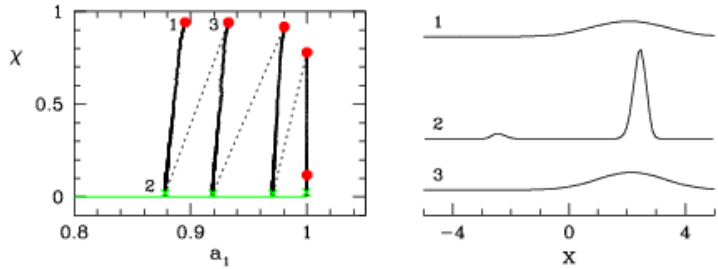}
}
\caption{Iterates of the moment map as seen in $a_1$, $\chi$ coordinates. Red
dots show points with Gaussian densities ${\hat \rho}(x;\bar{x}, \sigma)$.
The Gaussian densities corresponding to points labeled 1 and 3 are show to the
right. The evolution of the density over time $\tau$ is shown by bold curves
(actually fine series of bold points) with green crosses indicating the the
final time. The density corresponding to point 2 is shown to the right. Dashed
lines indicate the projections back to Gaussian density which preserve the
mean and standard deviation of the distribution.
}
\label{fig:an_enlarge}
\end{figure}

Figure \ref{fig:an_enlarge} shows the moment map in the coordinates suggested
by the preceding analysis.  The corresponding evolution of the density is
shown for one iterate of the map. We wish to view the evolution of the
density $\rho(x,t)$ in terms of the amplitudes $a_j(t)$ of projections onto
the eigenfunctions $\phi_j$.  We know that $a_0(t) \equiv 1$ so there is no
need to show this amplitude. The essential amplitude is $a_1$.  The effect of
all the higher amplitudes $a_j$, $j>1$ can be summarized by a single scalar 
$\chi$ defined as
$$ 
\chi \equiv 
\| \sum_{j=2}^{\infty}a_j\phi_{j} \|_2 =
\| \rho - a_0 \phi_0 - a_1 \phi_1 \|_2, 
$$
with $\rho=\rho(\cdot,\tau).$

\begin{figure}
\centerline{
\includegraphics[width=2in]{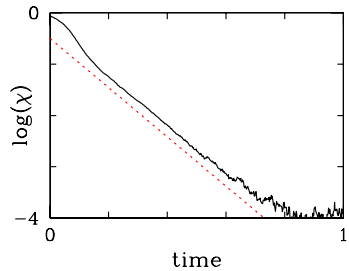}
}
\caption{Exponential decay of higher modes. Dashed curve has slope $-4.8$.  }
\label{fig:expplot}
\end{figure}

Consider point 1 in figure~\ref{fig:an_enlarge}.  To the right is shown the
Gaussian density $\rho(0) = {\hat \rho}(x;\bar{x}_0, \sigma_0)$ determined by
a point $(\bar{x}_0, \sigma_0)$ in moment space.  This density has a
significant projection onto the higher modes $\phi_j$, $j>1$ and $\chi$ is
significantly greater than zero.  From the analysis we expect the density to
evolve such that $\chi$ decays to zero on a time scale faster than the
dynamics of $a_1$. The thick black curves in \ref{fig:an_enlarge} show this
evolution.  There is little change in $a_1$ as the system evolves toward
$\chi=0$. The decay to $\chi=0$ is ultimately exponential as is shown in
figure~\ref{fig:expplot}. From this we estimate that $\lambda_2 \simeq
-5$. 

After time $\tau=1$ the system is at point 2 with corresponding density shown
at the right.  The resulting density is no longer Gaussian as expected from
\eqref{eq:approxm}.  The second-order moment map is obtained by determining
$(\bar{x}_1, \sigma_1)$, the mean and standard deviation of density 2.  For
the next iteration of the map one constructs a Gaussian density with this mean
and standard deviation, point 3 in Figure \ref{fig:an_enlarge}, and repeats.
The dashed line connecting points 2 and 3 illustrates this.  Thus the moment
map takes Gaussian ${\hat \rho}(x;\bar{x}_0, \sigma_0)$ determined by
$(\bar{x}_0, \sigma_0)$ (point 1) to ${\hat \rho}(x;\bar{x}_1, \sigma_1)$
determined by $(\bar{x}_1, \sigma_1)$ (point 3) where the map can be again be
iterated.  Note that the evolution from point 1 to point 2 is due to linear
flow of the Fokker-Planck equation. Nonlinearity results from the projection
of density at point 2 back to $(\bar{x}_1, \sigma_1)$, i.e. going from point 2
to point 3 introduces nonlinearity into the map.  Further note that the mean
has moved slightly to the right after one iteration of the map. Hence, on the
next iteration less of the mass will move into the left well, (as in point
2). In this way the map stabilizes the meta-stable states corresponding to
localization of density into a single well.

\begin{figure}
\centerline{
\includegraphics[width=2.5in]{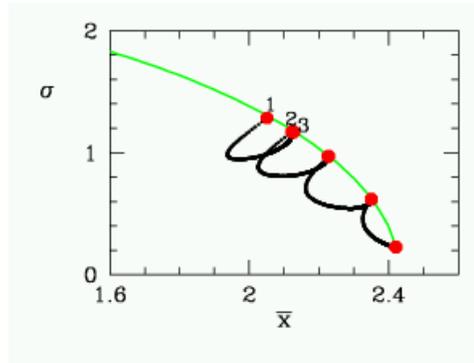}
}
\caption{Iterates of the moment map seen in $(\bar{x}, \sigma)$ coordinates.
The points correspond to exactly the same points as in Figure
\ref{fig:an_enlarge}.  The bold curves (actually fine series of bold points)
show the evolution of $(\bar{x}, \sigma)$ as the density evolves between
iterates of the map. The green curve is the slow manifold ($\chi=0$) also
shown in \ref{fig:an_enlarge}.
}
\label{fig:xs_enlarge}
\end{figure}

Now consider exactly the same evolution seen in the moment-map coordinates
$(\bar{x},\sigma)$ in Figure \ref{fig:xs_enlarge}.  The green curve is that
generated by $a_0=1$, $-1 \le a_1 \le 1$, $a_j=0$, $j>1$.  This curve is also
shown as green in Figure \ref{fig:an_enlarge}.  
This is the slow manifold for the
system. Starting from point $(\bar{x}_0,\sigma_0)$, point 1, the evolution of
$(\bar{x},\sigma)$ as $\rho(x,t)$ evolves is shown in bold.  This is the decay
of the modes $\phi_j$, $j>1$.  After time $\tau$ the system reaches
$(\bar{x}_1,\sigma_1)$, point 2.  Even though the transient dynamics is such
as to move initially away from the slow manifold, the evolution brings the
system back, as it must. The projection back to Gaussian density does
not change $\bar{x}$ and $\sigma$ so points 2 and 3 appear the same in
Figure \ref{fig:xs_enlarge}.

\begin{figure}
\centerline{
\includegraphics[width=2.5in]{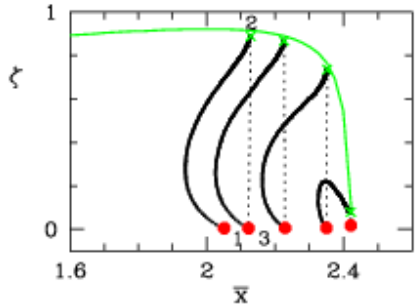}
}
\caption{ Iterates of the moment map seen in $(\bar{x}, \zeta)$ coordinates.
The points correspond to exactly the same points as in Figures
\ref{fig:an_enlarge} and \ref{fig:xs_enlarge}.  The bold curves
(actually fine series of bold points) show the evolution of $(\bar{x},
\zeta)$ as the density evolves between iterates of the map. The green curve
is the slow manifold.}
\label{fig:xz_enlarge}
\end{figure}

We know that as the system evolves from point 1 to point 2 the density became
non-Gaussian. We show this in Figure \ref{fig:xz_enlarge} by showing the
dynamics out of the $(\bar{x},\sigma)$ plane. Here $\zeta$ is given by
$$
\zeta \equiv \| \rho - \hat\rho(\cdot,\bar{x},\sigma) \|_2,
$$
with $\rho=\rho(\cdot,\tau)$
and is similar in spirit to $\chi$. $\zeta$ is a measure of how far
the density is from Gaussian whereas $\chi$ measure how far the density is
from the slow manifold.  However, these have opposite roles and behavior.  
Gaussian densities correspond to $\zeta=0$, but since such densities are 
not in general on the slow manifold they correspond to $\chi \ne 0$. 

Starting from point 1 in Figure \ref{fig:xz_enlarge}, $\zeta$ is necessarily
zero since the density (Figure \ref{fig:an_enlarge} point 1) is Gaussian. As
the system evolves toward the slow manifold and becomes non-Gaussian, $\zeta$
increases (point 2). Projection back to Gaussian moves the system vertically
downward to $\zeta=0$ (point 3).

Now consider the case $a_1 \simeq 1$, near the local minima of the potential
wells (we consider only the right well, the left is the same).  One sees in
Figures \ref{fig:an_enlarge} and \ref{fig:xz_enlarge} that points on the slow
manifold correspond to nearly Gaussian densities. This can be most clearly
seen in \ref{fig:xz_enlarge} where the slow manifold falls to near
$\zeta=0$. The end point of the green curve is $a_1=1$.  In Figure
\ref{fig:an_enlarge} we seen that the red points (the Gaussian densities in
$a$ coordinates) fall to very nearly $\chi=0$ meaning that these densities are
almost exactly represented by a sum of $\phi_0$ and $\phi_1$.  The small gap
between the slow manifold and the Gaussian density (seen in both Figures
\ref{fig:an_enlarge} and \ref{fig:xz_enlarge}) reflects that fact that the
metastable density is not exactly Gaussian.

\begin{figure}
\centerline{
\includegraphics[width=2.5in]{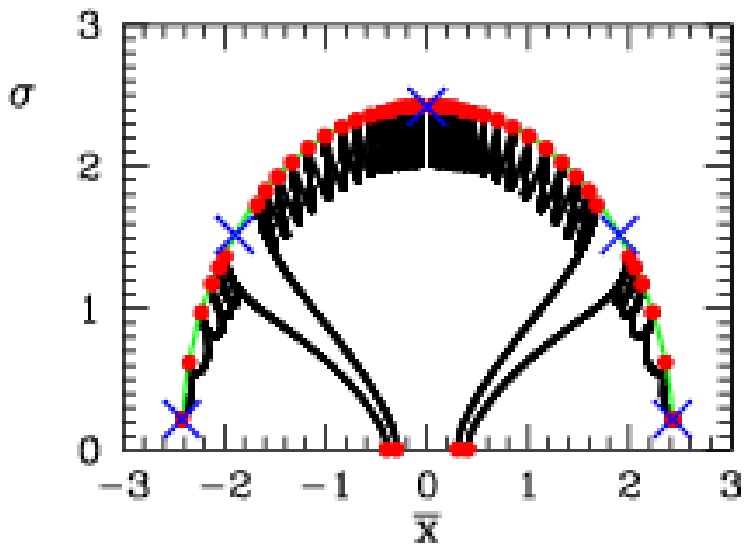}
}
\caption{ Iterates of the moment map seen in $(\bar{x}, \sigma)$
coordinates. Same as Figure \ref{fig:xs_enlarge} except over a larger range of
coordinates and same as \ref{fig:SDE1_pp_sym} except here $\tau=1$.  }
\label{fig:xs}
\end{figure}

In Figure \ref{fig:xs} we show the dynamics in $(\bar{x}, \sigma)$ coordinates
showing all the fixed point of the moment map. This is similar to Figure
\ref{fig:SDE1_pp_sym}(right) except that here $\tau=1$.  The figure shows
trajectories starting from four initial conditions.  Note that the slow
manifold accurately captures the unstable manifold of the saddle fixed points.

\subsection{2D SDE}

In this section we consider moment maps for the 2D SDE (\ref{eq:SDE2})
presented in \S\ref{ssec:2DSDE}.  The dynamics are potentially richer than for
the 1D SDE considered up to now. Nevertheless, the stabilization of metastable
states is the same as for the 1D SDE.

As before we examine maps for both first-order and second-order moments.  The
first-order map can be written as $\Phi(\bar{x}_1, \bar{x}_2)$ where
$\bar{x}_1$ and $\bar{x}_2$ are means of $Q$ and $P$ respectively.  The
second-order map can be written as $\Phi(\bar{x}_1, \bar{x}_2, \sigma_1,
\sigma_2, c)$ where $\sigma_1$ and $\sigma_2$ are the standard deviations of
$Q$ and $P$ respectively and $c$ is the cross correlation.  Other coordinates
could be used for the five-dimensional phase space.  We again consider the
slightly asymmetric double-well potential (\ref{eq:pot}) for parameters similar
to those use for the 1D SDE.  All results have been obtained numerically from
simulations of equations (\ref{eq:SDE2}) for the case $M=1$, $\gamma=1$,
$\beta=2$.

\begin{figure}
\centerline{
\includegraphics[width=4in]{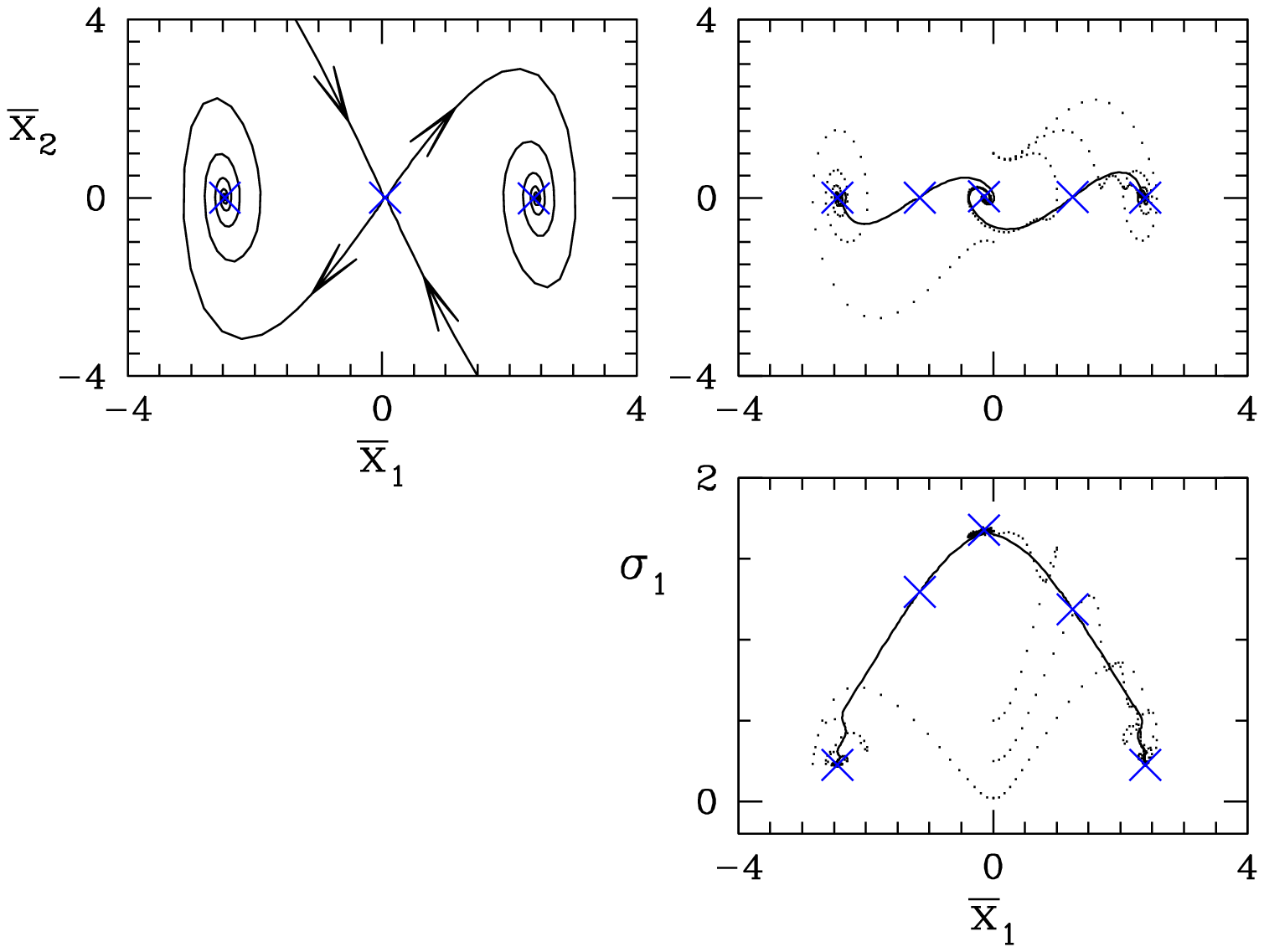}
}
\caption{Phase portrait for first-order (left) and second-order (right) moment
maps for the two-variable SDE with a slightly asymmetric double-well
potential.  Fixed points are indicated by crosses. For the first-order map,
stable and unstable manifolds of the saddle fixed point are shown. For the
second-order map, unstable manifolds of the saddle fixed point are shown. The
stable manifolds are four dimensional.  Points show some representative
trajectories.  Parameters are $\mu=6$, $\nu=0.3$, $\tau=0.1$.  }
\label{fig:SDE2_pp}
\end{figure}

Figure~\ref{fig:SDE2_pp} shows typical phase portraits for the first-order and
second-order moment maps. The first-order map exhibits the dynamics typical of
bistable damped oscillators.  Similar to the 1D case, the fixed points are
located at, or very close to, $(\bar{x}_c, 0)$ where $\bar{x}_c$ is a local
extremum of the potential.  The second-order map has five fixed points,
similar to those seen in figure~\ref{fig:SDE1_pp_sym}, including a stable
fixed point with a very broad Gaussian measure.  Typically trajectories of
interest approach either one or the other of the stable fixed points
corresponding to the potential minima.  The moment maps for the 2D SDE
stabilize metastable states in a very similar way to that found for the 1D SDE.

\begin{figure}
\centerline{
\includegraphics[width=3in]{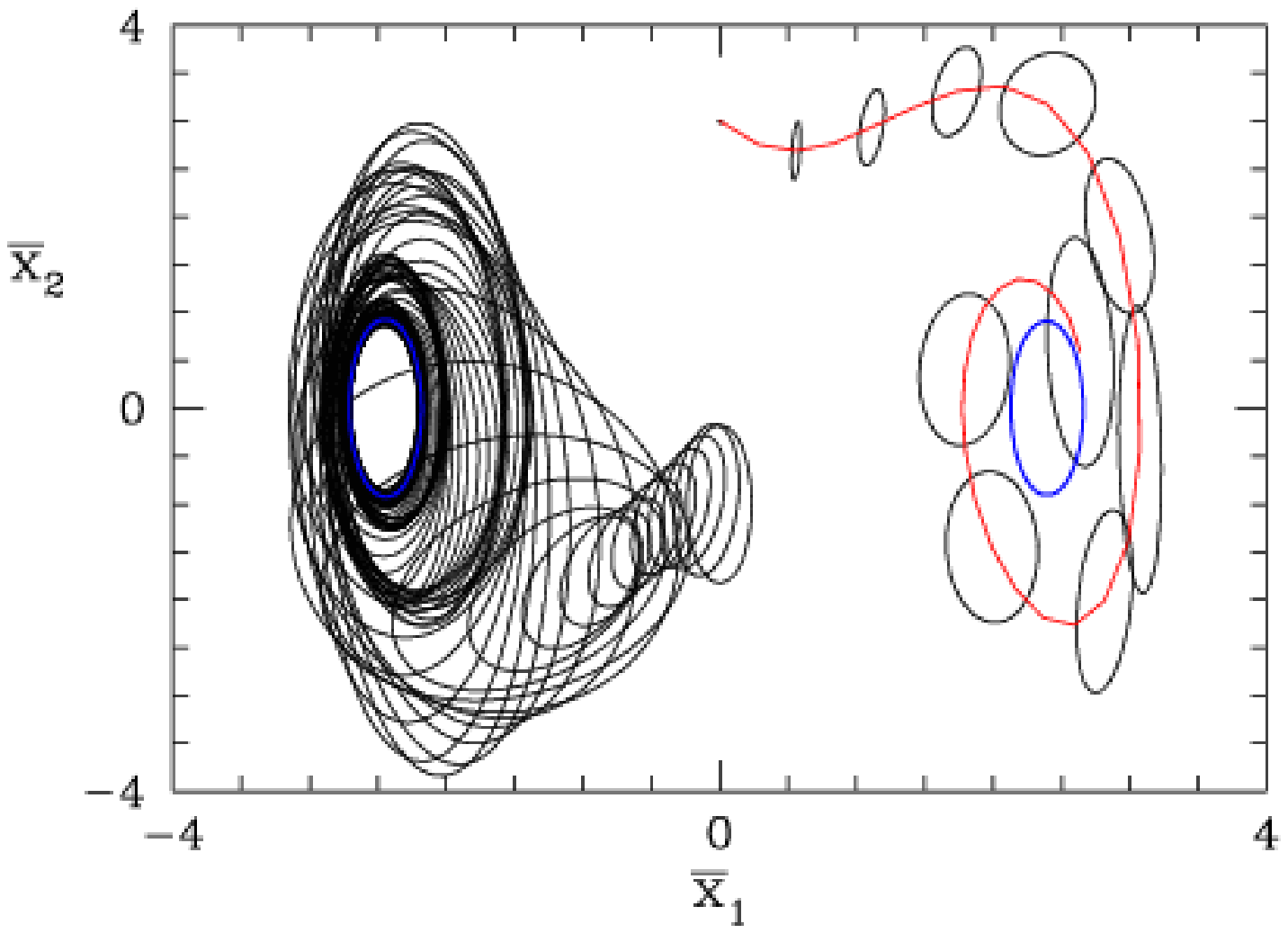}
}
\caption{Dynamics of the second-order moment map for the two-variable SDE. Two
  trajectories are visualized by plotting ellipses corresponding to points in
  the five-dimensional phase space. One trajectory starts at $(\bar{x}_1,
  \bar{x}_2, \sigma_1, \sigma_2, c) = (0, 3, 0.01, 0.01, 0)$. Only a few
  representative ellipses are shown.  The curve shows the path of the center
  of the ellipses.  The other trajectory starts at $(\bar{x}_1, \bar{x}_2,
  \sigma_1, \sigma_2, c) = (0, -1, 0.1, 0.5, 0)$. In this case every points on
  the trajectory is shown.  Two stable fixed points of the map are shown as
  bold blue ellipses. Parameters are $\mu=6$, $\nu=0.3$, $\tau=0.1$. 
}
\label{fig:SDE2_ellipse}
\end{figure}

Figure~\ref{fig:SDE2_ellipse} provides a better view of the dynamics of the
second-order moment map and shows how trajectories approach the stable fixed
points. A Gaussian measure in two variables can be visualized as an ellipse in
the plane specified by five numbers: the center, the semi-major and semi-minor
axes, and the orientation.  So for each point in the five-dimensional phase
space of the moment map we plot an ellipse in the plane. One can think of each
ellipse as corresponding to the level set of a Gaussian density.

Two trajectories are shown, one with the ellipse at every iteration of the map
plotted, and the other with only a few representative ellipses plotted. One can
see how densities evolve under the map toward stable equilibria.  Trajectories
from almost all initial condition distributions that are not too broad, 
(i.e., initialized within one of the two metastable wells, as discussed above)
evolve to one or
the other of the two fixed points similarly to what is shown in
figure~\ref{fig:SDE2_ellipse}.  

\begin{figure}
\centerline{ \includegraphics[width=4in]{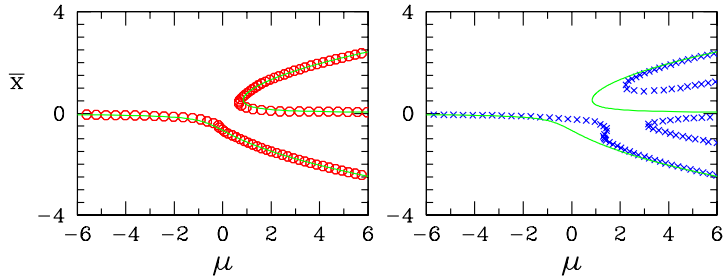} }
\caption{Bifurcation diagrams for the first-order (left) and second-order
  (right) moment maps for the two-variable SDE.  Parameters are $\nu=0.3$,
  $\tau=0.1$.  }
\label{fig:SDE2_bd}
\end{figure}

For completeness we show in figure~\ref{fig:SDE2_bd} bifurcation diagrams for
the moment maps for the 2D SDE.  These bifurcation diagrams reveal much the
same features as for the 1D SDE. Fixed points of the first-order moment map
track the potential minima.  For the second-order map, stable fixed points
corresponding to metastable states exist except in regions near where the
potential bifurcates from single- to double-well.  Clearly, close to such
parameter values, the separation of time scales between equilibration in one
well and transition to the other is no longer present, and the fixed points we
find depend on the map reporting horizon $\tau$.  Fixed points, both stable
and unstable, corresponding to broad distributions, separate the stable fixed
points.  While not shown, we find that the fixed points corresponding to the
stabilized metastable states are essentially independent of the time $\tau$
over which the map is defined; this suggests that the map is a good effective
description in their neighborhood, but not a useful one close to the
$\tau$-dependent fixed points.  The ability to test the sensitivity of the map
dynamics and fixed points to the parameter $\tau,$ as well as the ability to
use maps of different orders, allow potential tools with which to ``test
online" the validity of a given map as an effective model of the detailed
dynamics.

\subsection{Heat Bath}

Finally we consider the dynamics of a particle in a heat bath as described in
\S\ref{ssec:heat}.  Moment maps for this example are of the type defined for
ODE systems in \S\ref{ssec:ode}.  Nevertheless we expect the moment map for
the ODE system to behave very much like that of the 2D SDE since the SDE is
known to capture the dynamics of the heat bath model in the limit of a large
bath. From a computational viewpoint in fact there is not much distinction
between the stochastic (SDE) and the deterministic (ODE) cases since in both
case we numerically compute the moment maps using Monte Carlo simulations to
evolve densities over time interval $\tau$.

There are two related new issues, however. The first is that $N$, the number
of particles in the heat bath, is now a parameter which could potentially
affect the dynamics of the system.  The other related issue is that there is a
minimum time interval over which we should define the moment map.  For any
given bath size $N$ there is a maximum frequency $\omega_m$ of the oscillators
in the bath, where $\omega_m \simeq N^{1/3}$.  We therefore should take the
map time $\tau$ to be at least of the same order as the $1/\omega_m$.  We
always use $\tau > 2 \pi/\omega_m$.  

Other than the preceding two points the only significant difference between
the heat bath and the 2D SDE is that the heat bath requires significantly more
computation to evolve densities forward in time.  Hence the moment map is much
more expensive to compute for the heat bath than for the SDE.  For this
reason, we have limited our studies to a fixed potential: equation
\eqref{eq:pot} with $\mu=4$, $\nu=0.3$. This is the potential used for the
simulations shown in the introduction.

\begin{figure}
\centerline{
\includegraphics[width=4in]{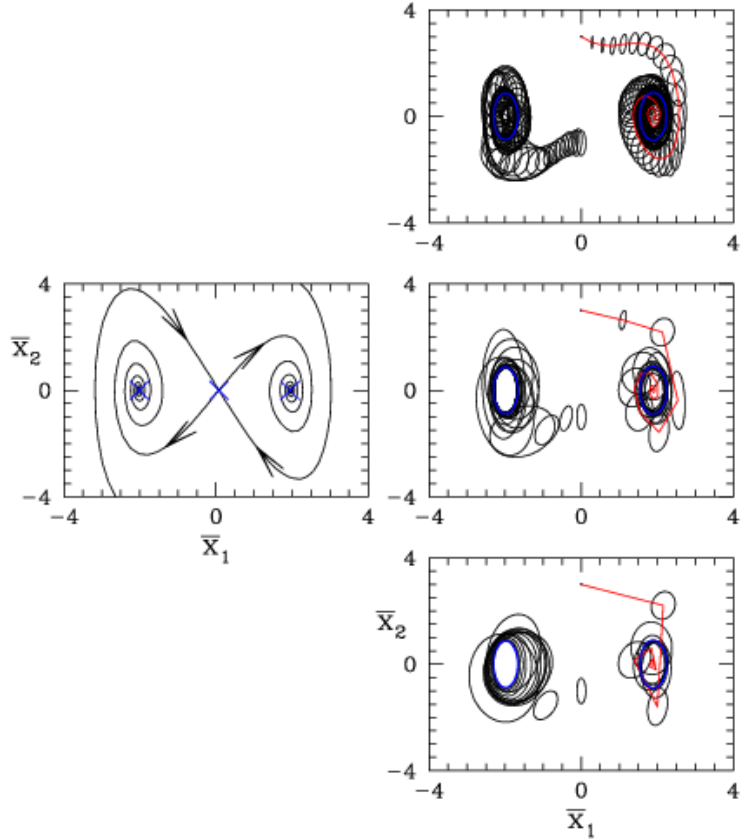}
}
\caption{Phase portraits showing the dynamics of moment maps for the heat
  bath model. First-order (left) and second-order (right) maps are shown. For
  the first-order map $N=8000$ particles is used. For the second-order map the
  following are used: (top) $N=256000$, $\tau=0.1$; (middle) $N=8000$,
  $\tau=0.4$; (bottom) $N=1000$, $\tau=0.8$.  }
\label{fig:bath_pp}
\end{figure}

Figure~\ref{fig:bath_pp} shows phase portraits of the moment map for the heat
bath model.  The first-order map shows the expected three fixed points at the
local extrema of the potential and exhibits the dynamics of bistable damped
oscillators. The map has indeed stabilized the metastable states with a well
defined boundary (the stable manifold of the saddle) separating the basins of
attraction.  The location of the fixed points does not depend in any
significant way on the number of particles in the bath or on the map time
$\tau$.  The stable and unstable manifolds do vary somewhat, primarily with
the map time $\tau$, but are always qualitatively as seen in
figure~\ref{fig:bath_pp}.

For the second-order map we show phase portraits in the style of
figure~\ref{fig:SDE2_ellipse} for the 2D SDE system. We show two trajectories
for each of three cases. The middle case is for $N=8000$, the value used for
simulations shown in the introduction.  The maximum oscillator frequency in
the bath is $\omega_m \simeq 20$ and so we use a map time of $\tau=0.4 >
\pi/10$.  

We show a map starting at $(\bar{x}_1, \bar{x}_2, \sigma_1, \sigma_2, c) = (0,
3, 0.01, 0.01, 0)$. This initial condition corresponds exactly to the initial
density used figure~\ref{fig:intro_ensemble}. One sees the similarity between
the trajectory for the moment map and the evolution of the density in
figure~\ref{fig:intro_ensemble}.  However, the ``leaking'' of mass into the
left potential well observed in figure~\ref{fig:intro_ensemble} is prevented
by the moment map via the mechanism illustrated in figure \ref{fig:an_enlarge}.
The moment map is nonlinear and has the two stable fixed points where the
density is only metastable. Most trajectories evolve to one or the other of
these fixed points.

For the same potential we have investigated the effect of $N$, the number of
particles in the heat bath, on the moment map.  We find that the map is very
insensitive to $N$ at least for $N \ge 1000$.  For large $N$ one can use a
smaller map time $\tau$ and the dynamics do depend weakly on $\tau$.
Specifically, the fixed points for the three cases in figure~\ref{fig:bath_pp}
are very slightly different, though this cannot be seen on the scale of the
figure. This difference is not due directly to $N$, but to the value of
$\tau$. 

\section{Conclusions}
\label{sec:conc}

In this paper we have introduced a mathematical framework intended to outline
and clarify some aspects of the coarse-grained approach to analyzing
stochastic and deterministic systems.  In particular we have given a precise
definition of the moment map.  These are maps on the (low-dimensional) space
of low-order moments of probability measures.  We have considered these maps
both for stochastic systems and for deterministic systems with random initial
data.  While the underlying evolution of densities in both systems is linear,
the moment maps are typically nonlinear. Our main focus has been understanding
the origins of this nonlinearity.

In this paper we sought to establish a connection between the dynamics of
coarse-grained observables (such as moments of evolving realization ensembles)
and the nonlinear dynamics one expects at the deterministic limit.
Contemporary estimation techniques would allow us to recover both the
deterministic and the stochastic component of an effective {\it stochastic}
model (e.g. \cite{Sahalia}).  Then, instead of using integral changes of
coarse-grained observables, we could directly seek the extrema of an
underlying effective potential, or even the extrema of the corresponding
equilibrium density (see, e.g. \cite{Mikko, Dima,Nanotubes}).

We have presented results for a number of model systems. We have first
considered the simple OU process for which a full analysis is possible.  Then
using a combination of numerical studies and analysis, we have considered in
most detail a one-dimensional SDE with a double-well potential. This system
provides the simplest example of a nonlinear moment map. In particular this
example shows how the moment map can stabilize metastable densities of the
underlying linear flow.  We have additionally presented numerical results for
moment maps computed for two other systems with double-well potentials - a
two-dimensional SDE and a deterministic ODE system with many degrees of
freedom.  Maps for both of these systems show the basic features found for the
one-dimensional SDE, namely nonlinearity and the stabilization of metastable
densities.

One of the important issue that arises naturally in this computational 
framework is the importance of {\it the observer}. How long does a physical
observer have to wait before declaring that an observed quantity is at 
steady state ? This is clearly related to our testing the dependence of the
map fixed points to the map reporting horizon. 
We also saw that the initialization of computational experiments (whether
within or not within a well) can be vital in the existence of an effective
reduced model (this is related, as we mentioned, to conditional averaging 
techniques).



{ \small
\begin{thebibliography}{99}

\bibitem{PNAS}
K. Theodoropoulos Y.-H. Qian and I. G. Kevrekidis
{\em ``Coarse" stability and bifurcation analysis using timesteppers:
	a reaction diffusion example}
Proc. Natl. Acad. Sci. USA {\bf 97}(18) (2000), pp.9840-9843.

\bibitem{GKT}
C. W. Gear, I. G. Kevrekidis and K. Theodoropoulos,  
{\em``Coarse" Integration/Bifurcation Analysis via Microscopic
	Simulators: micro-Galerkin methods}, 
Comp. Chem. Engng. {\bf 26} (2002) pp.941-963 

\bibitem{Manifesto}
I. G. Kevrekidis, C. W. Gear, J. M. Hyman, P. G. Kevrekidis, 
O. Runborg and K. Theodoropoulos,
{\em Equation-free coarse-grained multiscale computation: enabling microscopic
simulators to perform system-level tasks},
Comm. Math. Sciences {\bf 1}(4) (2003) pp.715-762; 
original version: physics/0209043 at arXiv.org.

\bibitem{ShortManifesto}
I. G. Kevrekidis, C. William Gear and G. Hummer,
{\em Equation-free: the computer-assisted analysis of complex, multiscale systems}
A.I.Ch.E Journal, {\bf 50}(7) (2004) pp.1346-1354.

\bibitem{Angel}
A. Angeletti, C. Castagnari, F. Zirilli
{\em Asymptotic eigenvalue degeneracy for a class of one--dimensional
     Fokker-Planck operators},
J. Math. Phys. {\bf 26}(1985), 678-691.


\bibitem{CHK00}
A. Chorin, O.H. Hald and R. Kupferman,
\newblock {\em Optimal prediction and the Mori-Zwanzig representation
of irreversible processes.}
\newblock Proc. Nat. Acad. Sci. USA {\bf97}(2000), 2968--2973.

\bibitem{DHF00}
P. Deuflhard, W. Huisinga, W. Fischer and C. Sch\"{u}tte,
\newblock {\em Identification of almost invariant aggregates in
reversible nearly uncoupled Markov chains.}
\newblock Lin. Alg. Appl. {\bf 315}(2000), 39--59.

\bibitem{ELvde04}
W. E, D. Liu and E. Vanden-Eijnden,
\newblock {\em Analysis of multiscale methods for stochastic differential 
equations}, Comm. Pure Appl. Math., (in press).

\bibitem{FK} G. W. Ford and M. Kac,
\newblock {\it On the quantum Langevin equation}.
\newblock J. Stat. Phys. {\bf 46} (1987), 803--810

\bibitem{Gardiner} C.W. Gardiner, 
\newblock {\it Handbook of Stochastic Methods}.
Springer-Verlag, Second Edition, 1985.

\bibitem{Huis} W. Huisinga, {\em Metastability of Markovian systems}.
PhD Thesis, Free University, Berlin, 2003.
{\tt http://page.mi.fu-berlin.de/~huisinga/publications/index.html}

\bibitem{GKS03}
D. Givon, R. Kupferman and A.M. Stuart,
\newblock {\em Extracting macroscopic dynamics: model problems and algorithms}.
\newblock Nonlinearity, {\bf 17}(2004), R55-R127. 

\bibitem{hal99} O. Hald,
\newblock {\em Optimal prediction of the Klein-Gordon equation}.
\newblock Proc. Nat. Acad. Sci. USA {\bf 96}(1999), 4774--4779.

\bibitem{KMC1}
A. Makeev, D. Maroudas and I. G. Kevrekidis, 
{\em "Coarse" stability and bifurcation analysis using stochastic
	simulators: Kinetic Monte Carlo Examples}
J. Chem. Phys. {\bf 116} (2002) 10083--10091.

\bibitem{KMC2}
A. G. Makeev, D. Maroudas, A. Z. Panagiotopoulos and I.G.Kevrekidis,
{\em Coarse bifurcation analysis of kinetic Monte Carlo simulations: a
	lattice gas model with lateral interactions},
J. Chem. Phys. {\bf 117}(18) (2002) pp.8229-8240.

\bibitem{KMC3}
A. G. Makeev and I.G. Kevrekidis, 
{\em Equation-Free multiscale computations for a lattice-gas model: coarse-grained
bifurcation analysis of the NO+CO reaction on Pt(100)},
Chem. Eng. Sci., {\bf 59}(8-9) (2004) pp.1733-1743.

\bibitem{Bubbles}
K. Theodoropoulos, K. Sankaranarayanan, S. Sundaresan and I.G.Kevrekidis, 
{\em Coarse Bifurcation Studies of Bubble Flow Lattice Boltzmann Simulations}
Chem.Eng.Sci., {\bf 59} (2004) pp.2357-2362; also nlin.PS/0111040 at arXiv.org.

\bibitem{Radek}
R. Erban, I. G. Kevrekidis and H. G. Othmer,
{\em An equation-free computational approach for extracting population-level
	behavior from individual-based models of biological dispersal}, 
      submitted to {\it Physica D}, Nov. 2004; also physics/0505179 at arXiv.org.

\bibitem{Graham}
C. Siettos, M. D. Graham and I. G. Kevrekidis,
{\em ``Coarse Brownian dynamics for nematic liquid crystals:
Bifurcation, projective integration and control via stochastic simulation}
J. Chem. Phys. {\bf 118}(22) (2003) pp.10149-10157 (2003); also
cond-mat/0211455 at arXiv.org.

\bibitem{Ehrenfest}
P. Ehrenfest,
{\em Collected Scientific Papers}, 
North-Holland, Amsterdam, 1959, pp. 213-300.

\bibitem{Gorban}
A. N. Gorban, I. V. Karlin, H. C. Oettinger, L.L. Tatarinova,
{\em Ehrenfest's argument extended to a formalism of nonequilibrium 
thermodynaics}
Phys. Rev. E, {\bf 63} (2001) 066124.

\bibitem{Kelley}
C. T. Kelley, I.G.K. and L. Qiao 
{\em Newton-Krylov Solvers for Time-Steppers},
submitted to SIADS (2004);  math.DS/0404374 at arXiv.org.

\bibitem{Kelleybook}
C. T. Kelley
{\it Iterative Methods for Linear and Nonlinear Equations}, 
SIAM Publiations, (1995).

\bibitem{HKRT94} C. Hillermeier, N. Kunstmann, B. Rabus and P. Tavan,
\newblock {\em Topological feature maps with self-organized
lateral connections: a population coded, one-layer model of associative memory},
\newblock {\em Biol. Cyber.}, {\bf 72}(1994), 103--117.

\bibitem{KSTT} R. Kupferman, A.M. Stuart, J. Terry and P. Tupper
\newblock {\em Long time behaviour of large mechanical systems with random
initial data}. 
\newblock Stochastics and Dynamics {\bf 2}(2002), 533-562.

\bibitem{Nelson} E. Nelson, {\em Dynamical Theories of Brownian Motion}. 
Available from {\tt http://www.math.princeton.edu/~nelson/books.html}


\bibitem{PS} G. Pavliotis and A.M. Stuart,
\newblock {\em White noise limits
for inertial particles in a random field}.
\newblock SIAM Multiscale Modeling and Simulation, {\bf 1}(2003), 527--553.

\bibitem{Doucet} A. Doucet, N. de Freitas and N.J. Gordon.
{\em Sequential Monte Carlo Methods in Practice}. Springer-Verlag (New York),
2001.

\bibitem{Schuette} C.Schuette, J. Walter, C. Hartmann and W. Huisinga
{\em An Averaging Principle for Fast Degrees of Freedom Exhibiting
Long-Term Correlations}. SIAM Multiscale Modeling and Simulation {\bf 2}(2004),
501-526.    


\bibitem{KevGear1}
C. W. Gear and I. G. Kevrekidis,
{\em Projective Methods for Stiff Differential Equations: problems with
gaps in their eigenvalue spectrum}
SIAM J. Sci. Comp. {\bf 24}(4) (2003) pp.1091-1106.

\bibitem{KevGear2}
C. W. Gear and I. G. Kevrekidis,
{\em Telescopic Projective Integrators for Stiff Differential Equations},
J. Comp. Phys. {\bf 187}(1) (2003) pp.95-109.  

\bibitem{KevGear3}
R. Rico-Martinez, C. W. Gear and I.G.Kevrekidis,
{\em Coarse Projective kMC Integration: Forward/Reverse Initial and Boundary 
Value Problems},
J. Comp. Phys., {\bf 196}(2) (2004) pp.474-489; nlin.CG/0307016 at arXiv.org.

\bibitem{GEAROLD}
C. W. Gear,
{\em Projective integration methods for distributions},
Technical Report
NEC TR 2001130, NEC, New Jersey, November (2001).

\bibitem{HUMMER1}
G. Hummer and I. G. Kevrekidis
{\em Coarse molecular dynamics of a peptide fragment: free energy, kinetics
and long time dynamics computations}
J. Chem. Phys. {\bf 118}(23) (2003) pp. 10762-10773; also physics/0212108.

\bibitem{SIMA}
S. Setayeshgar, C. W. Gear, H. G. Othmer and I. G. Kevrekidis
{\em Application of Coarse Integration to Bacterial Chemotaxis}
SIAM MMS {\bf 4}(1) (2005) pp.307-327 ; also physics/0308040.


\bibitem{LEVIN}
J. Cisternas, C. W. Gear, S. Levin and I.G.Kevrekidis,
{\em Equation-free modeling of evolving diseases: coarse-grained 
computations with
individual-based models},
Proc. Roy. Soc. London, {\bf 460}(2004) pp.27621-2779; also nlin.AO/0310011.

\bibitem{vde03}
E. Vanden Eijnden,
\newblock {\em Numerical techniques for multi-scale dynamical
systems with stochastic effects}.
\newblock Comm. Math. Sci. {\bf 1}(2003),
377--384.

\bibitem{Zw} R. Zwanzig,
\newblock {\it Nonlinear generalized Langevin equations}.
\newblock J. Stat. Phys. {\bf 9}(1973), 215--220.

\bibitem{Sahalia}
Y. Ait-Sahalia,
{\em Maximum-Likelihood Estimation of Discretely-Sampled Diffusions: 
A Closed-Form Approximation Approach}
Econometrica, {\bf 70} (2002) 223-262.

\bibitem{Mikko}
M. Haataja, D. Srolovitz and I.G.Kevrekidis, 
{\em Apparent hysteresis in a driven system with self-organized drag}, 
Phys. Rev. Lett. {\bf 92}(16) (2004) 160603; cond-mat/0310460 at arXiv.org.

\bibitem{Dima}
D. I. Kopelevich, A. Z. Panagiotopoulos and I. G. Kevrekidis,
{\em Coarse-Grained kinetic computations of rare events: application to 
micelle formation}, 
J. chem. Phys. {\bf 122} 044908 (2005); cond-mat/0407220 at arXiv.org.

\bibitem{Nanotubes}
S. Sriraman, I. G. Kevrekidis and G. Hummer, 
{\em Coarse nonlinear dynamics of filling-emptying transitions: 
water in carbon nanotubes} submitted to {\it Phys. Rev. Lett.}; cond-mat/0503491.

\end{thebibliography}
}
\end{document}